%% file: permeableVes.tex
\documentclass[9pt,twocolumn,twoside]{pnas-new2}
\templatetype{pnasresearcharticle} % Choose template 
\title{Hydrodynamics of a Semipermeable Vesicle Under Flow and
Confinement}

\author[a]{Bryan Quaife}
\author[a]{Ashley Gannon}
\author[b,1]{Y.-N.~Young} 

\affil[a]{Department of Scientific Computing, Florida State University, Tallahessee, FL 32306}
\affil[b]{Department of Mathematical Sciences, New Jersey Institute of
Technology, Newark, NJ 07102}

\leadauthor{Quaife} 

\authorcontributions{B.Q.~and Y.-N.Y.~designed and performed research
with A.G.'s assistance. B.Q.~and Y.-N.Y.~analyzed the data and wrote
the paper.}
\correspondingauthor{\textsuperscript{1} Corresponding author's e-mail: yyoung@njit.edu}


\begin{abstract}
Lipid bilayer membranes have a native (albeit small) permeability for
water molecules. Under an external load, 
provided that the bilayer structure stays intact and does
not suffer from poration or rupture,
a lipid membrane deforms and
its water influx/efflux is often assumed negligible in the absence of
osmolarity. 
In this work we use boundary integral simulations to investigate the
effects of water permeability on the vesicle hydrodynamics due to
a mechanical load, such as the viscous stress from an external flow
deforming a vesicle membrane in free space or pushing it through a
confinement. Incorporating the membrane permeability into the framework of Helfrich free energy for an inextensible, elastic membrane as a model for a semipermeable vesicle, we
illustrate that, in the absence of an osmotic stress gradient, the
semipermeable vesicle is affected by water influx/efflux over a
sufficiently long time or under a strong confinement. Our simulations
quantify the conditions for water permeation to be negligible in terms
of the time scales, flow strength, and confinement. These results shed
light on how microfluidic confinement can be utilized to estimate
membrane permeability.
\end{abstract}

\dates{This manuscript was compiled on \today}
\input my_macros2.tex

\setboolean{displaywatermark}{false}

\begin{document}

\maketitle
\thispagestyle{firststyle}
\ifthenelse{\boolean{shortarticle}}{\ifthenelse{\boolean{singlecolumn}}{\abscontentformatted}{\abscontent}}{}

\dropcap{W}ater exchange is essential for a living cell to adapt to its
environment over a wide range of time
scales~\cite{CadartVenkovaRechoEtAl2019_NaturePhys,
AlbertsMolecularBiology, YangMaVerkman2001_JBC,
SugieIntaglietta2018_AmJPhysiolHCP,
SaadounPapadopoulosWatanabeEtAl2005_JCS, Verkman2008_JMM,
BerthaudEtAl2016_SM, Keren2011_EurBJ, TaloniKardashSalmanEtAl2015_PRL}.  
%
%Water permeation across the cell membrane due to an osmotic shock has been well studied \cite{AlbertsMolecularBiology}. 
Recent findings confirm that both osmotic and mechanical stresses
contribute to significant water permeation for cells to migrate under
strong confinement~\cite{JiangSun2013_BJ, StrokaJiangChenEtAl2014_Cell,
LiMoriSun2015_PRL, yao-mor2017}, leading to a dynamic surface-to-volume
ratio of a migrating cell. While stress-induced release of messengers in
cells has been well-studied~\cite{Wan2008_PNAS, ForsythWan2011_PNAS, 
Russell-PuleriPazAdams2016_AJHCP, ZhangShenHoganBarakatMisbah2018_BJ,
GordonShimmelFrye2020_FP},
%osmotic effects on cellular dynamics have been well studied
%\cite{AlbertsMolecularBiology}, 
mechanically induced water permeation is often associated with membrane
poration or rupture under extreme
stresses~\cite{HarmanBertrandJoos2017_CJP,
RazizadehNikfarPaulLiu2020_BJ}. In this work we use modeling and direct
numerical simulations to show that, over the appropriate time scales or
under strong confinement, the intrinsic membrane permeability to water
can give rise to significantly distinct hydrodynamics of a semipermeable
vesicle without poration or rupture in the lipid bilayer membrane. 

\begin{figure*}[htp]
  \centering
  \includegraphics[width=0.9\textwidth]{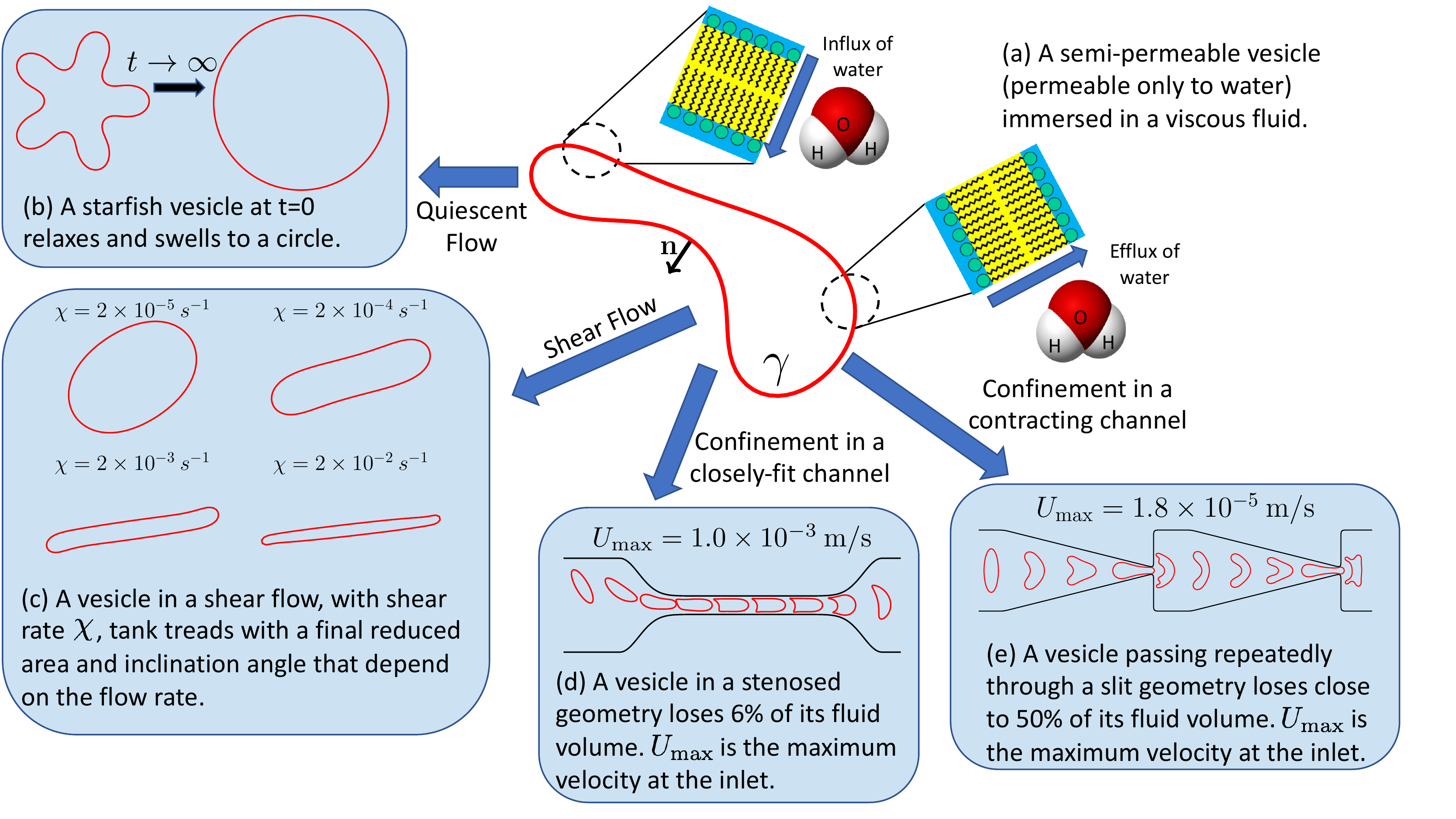}
  \caption{\label{fig:sketch} Schematics showing the effects of
  semipermeability on vesicle hydrodynamics in various configurations. 
  The lipid bilayer membrane (modeled as an interface $\gamma$ of zero
  thickness and an outward normal $\nn$) is permeable to water
  molecules, depending on the mechanical normal stress balance on the
  membrane.  The slipper in (a) is the steady-state shape of a
  semipermeable vesicle in a Poiseuille flow.}
\end{figure*}

The lipid bilayer membrane is permeable to water and lipid-soluble
molecules~\cite{Dick1964_JTB, FettiplaceHaydon1980_PhysRev,
DeamerBramhall1986_ChemPhysLipids, Grafmueller2019_ABLS}. 
%
%In the presence of an osmotic gradient across the membrane (such as a contrast in the concentration of an impermeable solute), water flows against the osmotic gradient  to neutralize the osmolarity. 
%
In the red blood cell membrane, aquaporin-1 channels give rise to a much
higher water permeability coefficient of approximately $1.8\times 10^{-2}$ cm/s in
erythrocytes~\cite{YangMaVerkman2001_JBC} while the intrinsic water
permeability of a lipid bilayer membrane is in the range of
$10^{-4}$--$10^{-3}$ cm/s~\cite{ThompsonHuang1966_ANYAS,
FettiplaceHaydon1980_PhysRev, Grafmueller2019_ABLS, Dimova2020_GVB,
BhatiaRobinsonDimova2020_SoftMatt}. 
Without aquaporins or other membrane channels that promote water
permeability, the intrinsic membrane permeability is known to depend on the
temperature, lipid composition~\cite{OlbrichRawiczNeedhamEtAl2000_BJ},
and the hydration of the membrane~\cite{MarrinkBerendsen1994_JPhysChem}.
%
%The swelling/shrinking of a cell is inevitably coupled to changes in
%membrane tension and deformation. Under a strong confinement (such as
%going through a microcapillary or submicron slit),  the transmembrane
%water flow may have comparable contributions from both osmotic and
%mechanical stresses
%\cite{StrokaJiangChenEtAl2014_Cell,LiMoriSun2015_PRL,yao-mor2017}.  
Molecular dynamic (MD) simulations show that a stretched red blood cell
membrane porates when the membrane tension reaches the order of $2$--$4$
mN/m (\cite{RazizadehNikfarPaulLiu2020_BJ} and references therein),
giving rise to enhanced membrane permeability to both water and
macromolecules. Consequently, as long as the lipid bilayer membrane
remains intact (no poration or rupture), the total amount of water inside the membrane 
is assumed constant throughout the experiment of duration
no longer than tens of minutes.
%
%membrane poration is inevitable when membrane tension is of the order
%of $2-4$ mN/m (\cite{RazizadehNikfarPaulLiu2020_BJ} and references
%therein). 
%%The CGMD results further show that a stretched RBC membrane may have pores that might contribute to transmembrane transport of both water and ATP molecules \cite{RazizadehNikfarPaulLiu2020_BJ}. 
%Thus an important outstanding question in stress-induced ATP release of
%highly deformed RBCs is if there is any transmembrane flow when RBCs
%release ATP as they experience large mechanical load due to confinement
%or large shear stress?  To address such question, the first step we
%take is to focus on the permeable water flux induced by mechanical
%stress.
%%
%%RBCs are known to release ATP under large deformation and microfluidic experiments have been used to study the correlation between RBC deformation and ATP release \cite{Wan2008_PNAS,ForsythWan2011_PNAS,ZhangShenHoganBarakatMisbah2018_BJ}.
%%
%%Why is it interesting to study the dynamics of a semipermeable membrane induced by mechanical stress/confinement?
%%
%%The osmolarity-induced water flux has been well studied, while mechanical pressure may also contribute to the permeable water flow \cite{yao-mor2017}.
%%The response of a cell to an osmotic gradient is inevitably both mechanical (tension redistribution and shape change) and chemical (water flux across the membrane).  
Under what condition is this a good assumption? Do liposomes conserve
volume under flow over a long time (hours)? What is the dynamic consequence when water
influx/efflux affects the membrane hydrodynamics? To answer these
questions, we investigate the balance between hydrodynamics stresses,
tension, and elastic stress of a semipermeable membrane under various
flowing conditions and confinement (Figure~\ref{fig:sketch}).
%Specifically our numerical simulations yield the quantification of the
%time scale and length scale over which water permeation becomes
%important. 
%
%
% In this work we use numerical simulations to examine the hydrodynamics
% of a semipermeable vesicle in two dimensions. 

Vesicles 
%(self-enclosed lipid bilayer membranes) 
have been used as a
model system to study reshaping, remodeling, and scission of cell
membranes due to osmotic stress~\cite{OgleckaEtAl2014_eLife, BhatiaRobinsonDimova2020_SoftMatt,
CamposSaric2020_bioRxiv, Dimova2020_GVB, BhatiaChrist2020_SoftMatt}.
%Under an osmotic shock, a vesicle may undergo oscillatory variation in
%its enclosed volume~\cite{ChabanonHoLiedberg2017_BJ}. 
%When trapped in a
%microfluidic channel, an osmotic shock induces vesicle shape changes,
%tank-treading and tumbling
%motions~\cite{BhatiaRobinsonDimova2020_SoftMatt,
%BhatiaChrist2020_SoftMatt}, an 
These results illustrate that the permeating water
flow is inevitably coupled with the mechanical pressure jump across the
membrane~\cite{yao-mor2017}, which could become comparable to osmotic
stress when the membrane deformation is large~\cite{LiMoriSun2015_PRL,
yao-mor2017}. 
In microfluidic experiments, it is challenging to keep a freely
suspended red blood cell or vesicle in a steady flow for more than a few
hours. It is also difficult to quantify the vesicle surface-to-volume
ratio accurately~\cite{MinetttiCallensCoupier2008_AppliedOptics}. Using
numerical simulations we can quantify the effects of water permeability
on the hydrodynamics of a vesicle under mechanical stresses.  Without an
osmotic gradient, we first show that a freely suspended semipermeable
vesicle behaves differently from an impermeable vesicle over a long
time. Next we show that
%, at normal physiological conditions,
the permeating water flow can be amplified by extreme confinement to give rise to observable change in both hydrodynamics and water content. 
These results provide insight into quantifying the effects of membrane permeability on vesicle hydrodynamics in microfluidics.
%
%shed light on the dynamic aging process of
%living cells that constantly go through flows and constrictions over a
%period of days.
 
%%%%%%%%%%%%%%%%%%%%%%%%%%%%%%%%%%%%%%%%%%%%%%%%%%%%%%%%%%%%%%%%%%%%%%%%
\section*{Formulation}
\begin{figure*}[htp]
  \centering
  \includegraphics{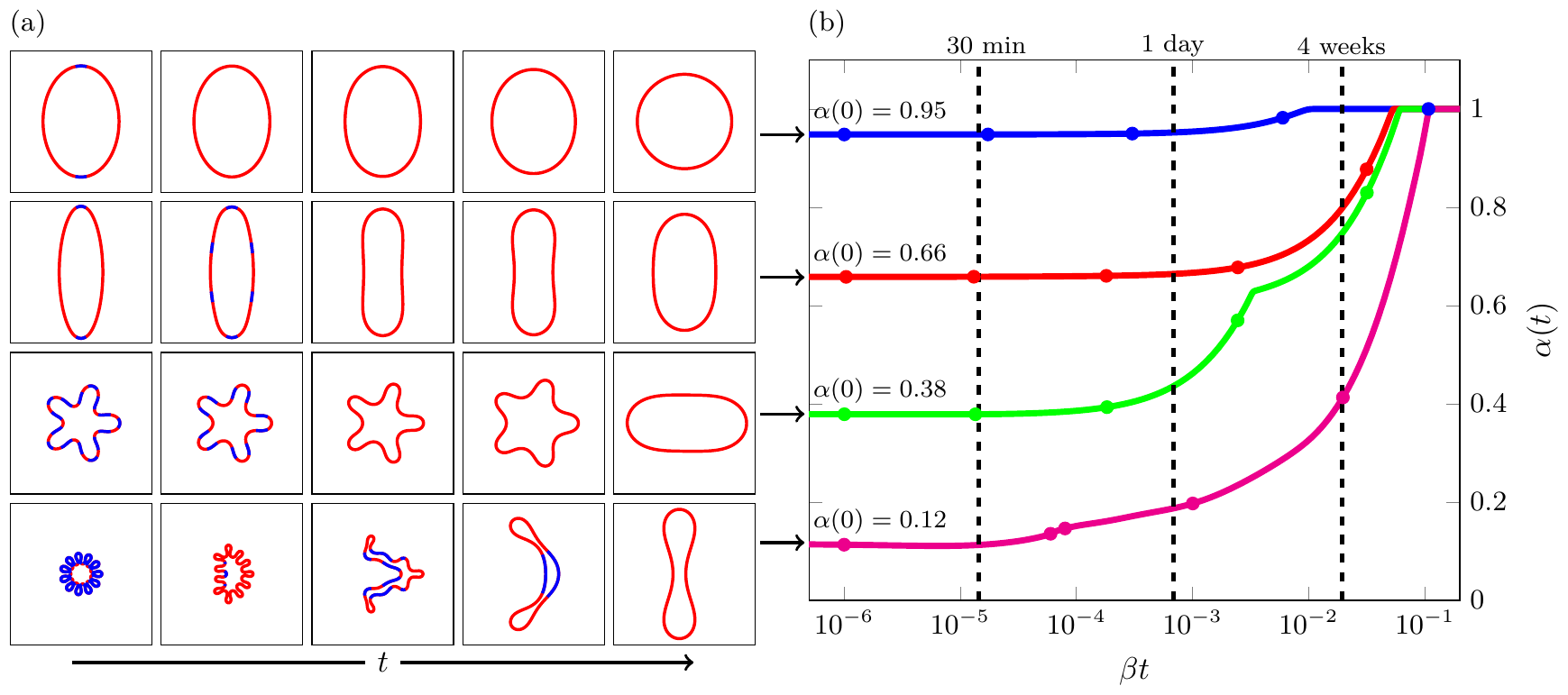}
  \caption{\label{fig:relaxationComposite} (a) Snapshots of four
  semipermeable vesicles with $\beta = 10^{-8}$ submerged in a fluid
  with no background flow. Time goes from left to right. The red regions
  correspond to influx and the blue regions correspond to efflux. In all
  cases, the vesicle reaches a steady state circular shape. (b) The
  reduced area of each of the simulations. The particular snapshots in
  (a) occur at the marks along the curve.}
\end{figure*}
We consider a two-dimensional semipermeable vesicle (permeable only to
water) suspended in a viscous fluid (Figure~\ref{fig:sketch}), and we
let $\gamma$ denote the membrane boundary. We let
$\alpha=4\pi A/L^2$ be the vesicle's reduced area where $A$ is its area
and $L$ is its length. To focus on the effect of water flow on vesicle
hydrodynamics in various conditions, we assume the fluid is the same
inside and outside the vesicle membrane. On the membrane the permeable
water flux gives rise to a difference between the fluid velocity
${\uu}({\xx}\in \gamma)$ and the membrane velocity
$\dot{\xx}$~\cite{yao-mor2017}:
\begin{align}
\label{eq:water_flux}
  \uu - \dot{\xx} = - k_w (R T \triangle c + \ff_\mathrm{mem} \cdot \nn) \nn, \qquad
  \xx \in \gamma,
\end{align}
where $\nn$ is the outward normal (Figure~\ref{fig:sketch}), $k_w$ is
the hydraulic conductivity (m$^2$s/kg), $R$ is the ideal gas constant,
$T$ is the temperature (K), $\triangle c$ is contrast in a solute
concentration (mol/volume), and $\ff_\mathrm{mem}$ is the membrane
stress that consists of both bending $ \ff_\mathrm{ben}= -k_b
\xx_{ssss}$ and tension $\ff_\mathrm{ten}=(\sigma \xx_s)_s$, where $s$
is the arclength, $\sigma = \Lambda - \frac{3}{2}\kappa^2$, $\Lambda$ is the
membrane tension, and $\kappa$ is the membrane curvature (see
Supplementary Material). Using $k_w\sim 10^{-12}$ m$^2$s/kg 
%(typical of mammalian cell membranes~\cite{LiMoriSun2015_PRL}) 
and an osmolarity $\triangle c = 0.1$ mol/L (consistent with the osmotic
filtration experiments~\cite{OlbrichRawiczNeedhamEtAl2000_BJ}), the
corresponding water flow is of the order $v_\mathrm{o}=0.1 \; \mu$m/s.
Assuming a membrane bending stiffness of $k_b = 10^{-19}$ J, a membrane
tension of $10^{-3}$ N/m (one tenth of the lysis tension), and a
characteristic length scale of $10^{-6}$ m, the tension contribution
($v_\mathrm{t}$) and elastic stress contribution ($v_\mathrm{e}$) to the
permeable water flow can be scaled to $v_\mathrm{o}$ as
$v_\mathrm{o}:v_\mathrm{t}:v_\mathrm{e}=10^{0}:10^{-2}:10^{-6}$.
Furthermore, we note that in the absence of an osmotic stress
($\triangle c = 0$), the permeating water flow may be non-negligible
when the membrane tension becomes large due to the external stress such
as the hydrodynamic stress from an external flow or confinement. In the
following we explore both.

Specifically we focus on the mechanical component of the water flux
in~\eqref{eq:water_flux} with $\triangle c =0$. In our
non-dimensionalization the hydraulic conductivity is scaled to $R_0/\mu$
(where $R_0$ is a characteristic length and $\mu$ is the solvent
viscosity): $\beta = k_w/(R_0/\mu)$ (see Supplementary Material),
$\beta=0$ for an impermeable vesicle, and $\beta>0$ for a semipermeable
vesicle. Under strong confinement, a large value of $\beta$ is needed in
our simulations to keep the membrane tension below the threshold value
for membrane poration $\sim 10^{-3}$ N/m. Such large value of $\beta$
may pertain to red blood cell membranes with aquaporin proteins.

%%%%%%%%%%%%%%%%%%%%%%%%%%%%%%%%%%%%%%%%%%%%%%%%%%%%%%%%%%%%%%%%%%%%%%%%
\section*{Hydrodynamics of a semipermeable vesicle in free space}
To highlight the effects of membrane permeability to water, we consider
a single semipermeable vesicle, with (dimensionless) permeability
coefficient $\beta>0$, in familiar configurations: In free space we
place a semipermeable vesicle in a quiescent flow, a planar shear flow,
and a Poiseuille flow. For each example, we fix the vesicle length and
examine the enclosed water content as a dynamic consequence of water
flow, vesicle shape deformation, and membrane tension distribution.
Under these flows, our simulations show
that the steady state of an impermeable vesicle is different from that of a semipermeable vesicle, which depends only on the strength of the external flow and is insensitive to permeability $\beta$ (as long as $\beta>0$). We find that  the time in the dynamic evolution towards equilibrium is scaled by $\beta^{-1}$: the smaller $\beta$ is the longer it takes to reach the steady state.

In experiments, vesicles under shear flows are measured for a few
minutes (chapter 19 in~\cite{Dimova2020_GVB}), a duration over which a
vesicle can be assumed impermeable~\cite{AbkarianViallat2005_BJ}. Our
simulations of a semipermeable vesicle under these flows in open
geometry show that water permeability may lead to different vesicle
hydrodynamics only after a duration of $\sim 1$ day.

%short time scales.

%%%%%%%%%%%%%%%%%%%%%%%%%%%%%%%%%%%%%%%%%%%%%%%%%%%%%%%%%%%%%%%%%%%%%%%%
\subsection*{Quiescent Flow} 
\begin{figure*}[htp]
  \centering
  \includegraphics{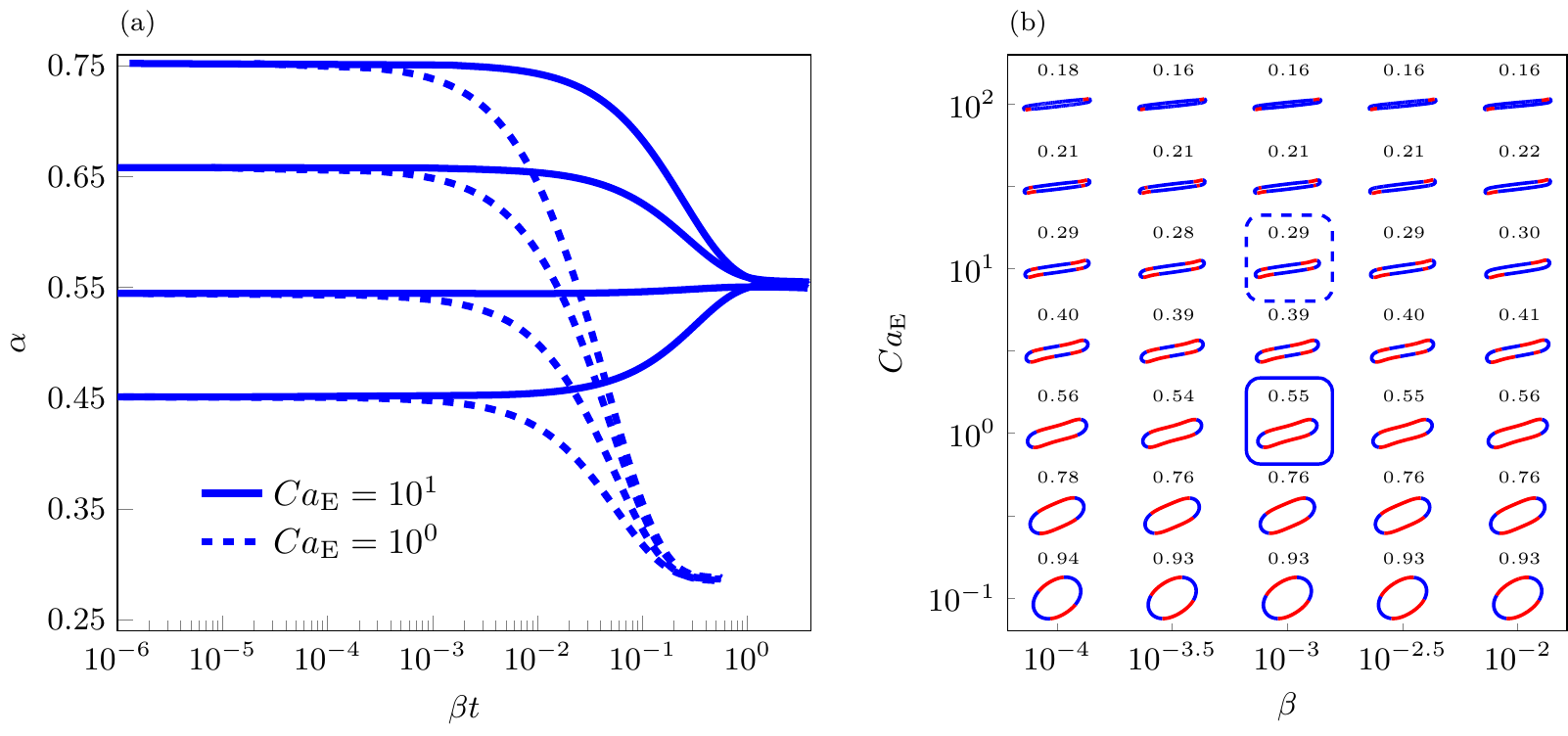}
  \caption{\label{fig:shearComposite} (a) The reduced area of a
  semipermeable vesicle with $\beta = 10^{-3}$ initialized with four
  different reduced areas in a shear flow with two different flow rates.
  (b) The equilibrium shape of a vesicle with varying flow rates and
  membrane permeability. The circled vesicles correspond to the
  simulations in part (a).}
\end{figure*}
In the absence of a background flow, a semipermeable vesicle in free
space can be quantified in terms of its area dynamics and its transition
to the equilibrium shape. Our analysis in Supplementary Material shows
that a semipermeable vesicle relaxes to a circle at a rate that is
proportional to $\beta$, and the steady state tension is $\Lambda =
2\pi^2/L^2$. Our analysis further shows that the time for a
semipermeable vesicle to deviate from its initial reduced area and to
reach its steady state is proportional to $\beta^{-1}$. For this reason,
our numerical results are reported in terms of the scaled time $\beta
t$.

In Figure~\ref{fig:relaxationComposite}, we examine the relaxation
dynamics of a semipermeable vesicle with four initial shapes and reduced
areas as labeled. The vesicles on the left are color-coded by the water
permeation flux: red denotes an influx into the vesicle, and blue
denotes an efflux out of the vesicle. First we observe that the initial
vesicle shape determines the direction of water flux. For the vesicle
with an initial reduced area $\alpha(0)=0.12$ (bottom row) the vesicle
deflates (blue regions) first due to the high interior capillary
pressure that corresponds to the high membrane curvatures. For the other
three cases efflux is dominated by influx, especially after $t=1$ day
when the increase in total water content becomes non-negligible.
%
%is found at some locations of high curvature, while
%mostly water permeates into the vesicle and significant change in the
%reduced area is found at $t\sim 1$ hour. 
%
Over time (about four weeks) all vesicles inflate and as expected, each
vesicle eventually reaches a circular shape with reduced area
$\alpha=1$. 
%%These regions are denoted in blue and are only present at early times. 
%Finally, the vesicles's reduced area do not steady state tension is
%$\Lambda = 2\pi^2/L^2$. 
%Therefore, the times for the vesicle to deviate from its initial reduced area and to reach its
%steady state are proportional to $\beta^{-1}$. 
%
%Each of the four cases in Figure~\ref{fig:relaxationComposite} eventually
%reaches a circular shape with reduced area $\alpha=1$. Next, we observe
%that even though the vesicle is inflating to a circle, 
%%meaning that there is a net flux into the vesicle, 
%there are regions where the flux
%is out of the vesicle. These regions are denoted in blue and are only
%present at early times. 
%
%Finally, the vesicles's reduced area do not
%change by much at early times (up to ten minutes), and 

Between $t=30$ minutes and $t=1$ day, the total water content is nearly constant.
%
%the intermediate shapes of
%semipermeable vesicles resemble the shape of an impermeable vesicle.
%
This can be explained by considering the two terms that contribute to
the vesicle velocity: (i) the permeability velocity, $\beta(\ff \cdot
\nn) \nn$, and (ii) the force balance velocity, $\SS[\ff]$. Since
$\beta\ll 1$, the permeability velocity is much smaller than the force
balance velocity at early times, and the semipermeable vesicle dynamics
resemble those of an impermeable vesicle. As the force balance velocity
decreases, the permeability velocity dominates, and the vesicle begins
to inflate to a circle with vanishing force and velocity.

%%%%%%%%%%%%%%%%%%%%%%%%%%%%%%%%%%%%%%%%%%%%%%%%%%%%%%%%%%%%%%%%%%%%%%%%
\subsection*{Planar Shear Flow}
We consider a semipermeable vesicle in a planar shear flow,
characterized by the elastic capillary number $Ca_\mathrm{E}=\chi\tau$ where $\tau$ is the characteristic time scale from balancing the viscous stress with the elastic stress
%in the non-dimensionalization 
(see Supplementary Material) and $\chi$ is the shear rate of the planar
shear flow $\uu_{\infty}(\xx) = \chi (y,0)$.  When $\beta=0$
(impermeable case), the vesicle reduced area and fluid viscosity
contrast determine whether a vesicle tank treads or
tumbles~\cite{fin-lam-sei-gom2008, kra-win-sei-lip1996}. Focusing on
semipermeable vesicles with no viscosity contrast, we find that a
semipermeable vesicle tilts to an inclination angle and undergoes
tank-treading dynamics, similar to the case of an impermeable vesicle.

In Figure~\ref{fig:shearComposite}(a), using four initial reduced areas
and two flow rates, we plot the reduced area of a semipermeable vesicle
with $\beta = 10^{-3}$. As observed in a quiescent flow, the final
vesicle shape is independent of the initial reduced area. However, it
does depend on the flow rate. We also observe that the amount of water
inside the vesicle remains constant until $t\sim 2$ hours, when the
normal component of the flow velocity becomes significantly smaller than
the permeable velocity, leading to a changing reduced area.

For a variety of dimensionless permeability coefficient $\beta>0$ and
capillary number $Ca_\mathrm{E}$, a semipermeable vesicle reaches an
equilibrium shape as in Figure~\ref{fig:shearComposite}(a). Since the
equilibrium shape depends on $Ca_\mathrm{E}$ and is independent of the
initial reduced area, we initialize the vesicle with reduced area
$\alpha = 0.65$ for all simulations in
Figure~\ref{fig:shearComposite}(b), where we summarize the equilibrium
vesicle shape and its reduced area. We see that the capillary number
(flow rate) significantly affects the final vesicle shape, but the
permeability coefficient has a very minor effect. Instead, as we saw in
the quiescent analysis, $\beta$ sets a time scale since the time
required for a vesicle to reach its steady state configuration is
proportional to $\beta^{-1}$.

%%%%%%%%%%%%%%%%%%%%%%%%%%%%%%%%%%%%%%%%%%%%%%%%%%%%%%%%%%%%%%%%%%%%%%%%
\subsection*{Poiseuille Flow}
\begin{figure*}[htp]
  \centering
  \includegraphics{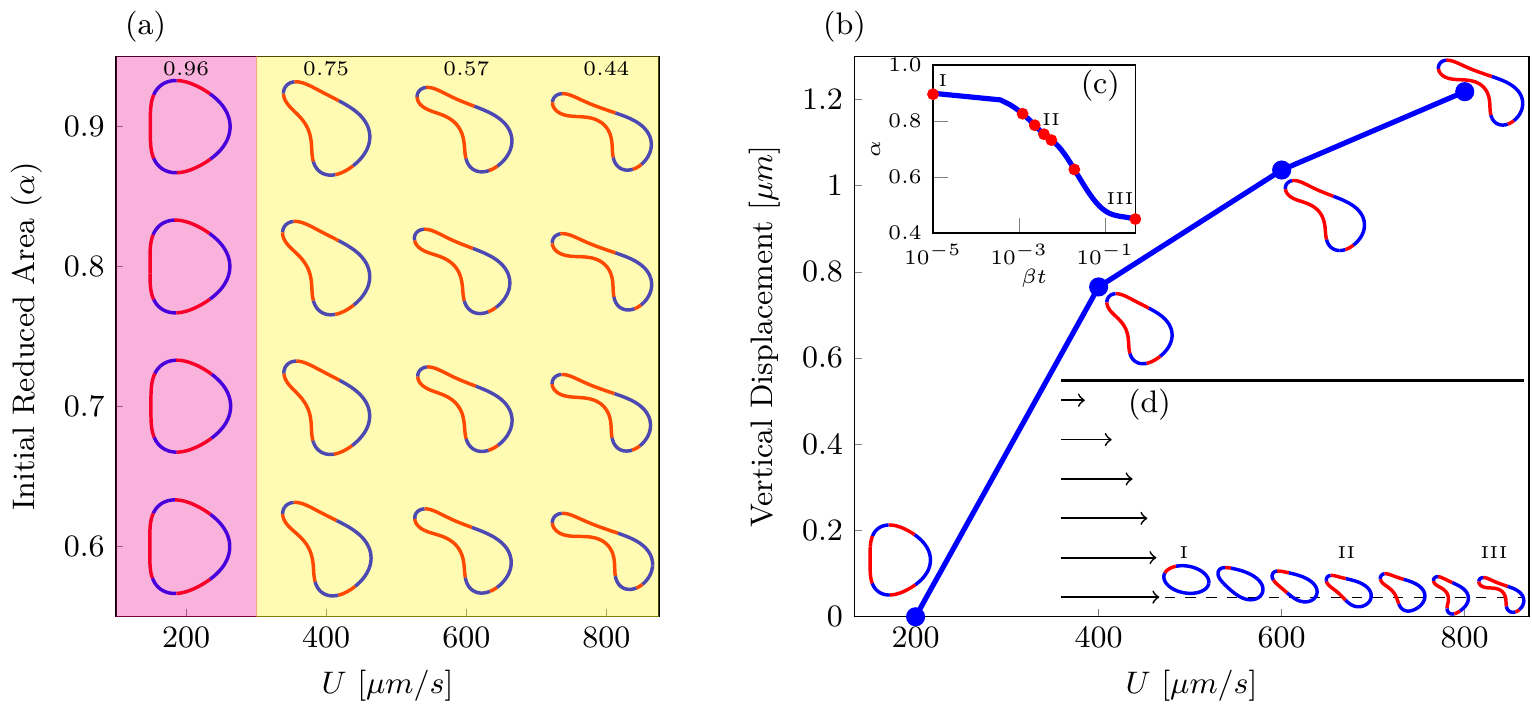}
  \caption{\label{fig:parabolicComposite} (a) The equilibrium shape of
  a semipermeable vesicle submerged in a Poiseuille flow with varying
  initial reduced areas and flow rates. The red regions correspond to
  influx and the blue regions correspond to efflux. (b) The steady
  state vertical displacement at four different flow velocities. The
  steady state shapes are superimposed. (c) The reduced area of the
  slipper formed with the flow rate of $800 \; \mu$m/s. (d) The vesicle
  shape shown with the background imposed flow. The channel width is
  12.5 times larger than the vesicle radius. The corresponding reduced
  area of each vesicle is indicated by the marks in plot (c).}
\end{figure*}

We consider a semipermeable vesicle in the Poiseuille flow $\uu =
U(1-(y/W)^2,0)$, where $U$ is the maximum velocity at $y=0$. The
non-linear effects of the Poiseuille flow on the migration of an
impermeable vesicle have been well-studied~\cite{kao-bir-mis2009}: At a
given value of the maximum velocity $U$, there exists a critical reduced
volume above which the equilibrium vesicle shape is symmetric (parachute
or bullet shapes) and the vesicle stays at $y=0$. Below this critical
reduced volume the equilibrium vesicle moves off the center and is
asymmetric (tank-treading slipper shape). We validated our numerical
codes by comparing against results in Kaoui {\em et
al.}~\cite{kao-bir-mis2009} (see Supplementary Material).

%
%
%We examine the effect of the non-linear Poiseuille flow on the migration of
%semipermeable vesicle. In this experiment, the vesicle is initialized
%above the center line $y=0$, and it migrates towards the center
%line~\cite{dan-vla-mis2009}. 
%
%For an impermeable vesicle, depending on
%the vesicles reduced area and the flow rate, the equilibrium vesicle shape
%is either an axisymmetric parachute or bullet, or an asymmetric
%tank-treading slipper~\cite{kao-bir-mis2009}. 
%We validated our numerical codes by comparing against results in Kaoui {\it et al.} \cite{kao-bir-mis2009}.
%One of the most salient features of an impermeable vesicle in the Poiseuille flow is that, 
%%
%%To validate our methods,
%%in Figure~\ref{fig:parabolicComposite}(a), we construct a phase diagram
%%for the steady state shape of an impermeable vesicle
%%vesicle~\cite{kao-bir-mis2009} (see Figure 2). While the vesicle shapes
%%qualitatively agree, Kaoui et al.~report that the vesicles with
%%$\nu=0.7$ and flow rates $U = 1200 \mu m/s$ and $U = 1600 \mu m/s$ are
%%axisymmetric. In our simulations these vesicles are slightly asymmetric,
%%are undergoing a tank treading motion, and we therefore report them as
%%asymmetric slippers.
%
%Having validated our method, we investigate the dynamics of a semipermeable vesicle in a Poiseuille flow. 
%
%Using several initial
%$U(1-(y/W)^2,0)$, where $U$ is the maximum imposed velocity. We examine
%the effect of the non-linear Poiseuille flow on the migration of
%semipermeable vesicle. 
%

At the beginning of the simulations, a semipermeable vesicle is placed
above the center line $y=0$ and it migrates towards the center line as
in the impermeable case~\cite{dan-vla-mis2009}. 
%
%However,
%for an impermeable vesicle, depending on the initial reduced area and
%the maximum flow velocity $U$, the equilibrium vesicle shape is either
%an axisymmetric parachute or bullet, or an asymmetric tank-treading
%slipper~\cite{kao-bir-mis2009}. 
%
%To validate our methods,
%in Figure~\ref{fig:parabolicComposite}(a), we construct a phase diagram
%for the steady state shape of an impermeable vesicle
%vesicle~\cite{kao-bir-mis2009} (see Figure 2). While the vesicle shapes
%qualitatively agree, Kaoui et al.~report that the vesicles with
%$\alpha=0.7$ and flow rates $U = 1200 \mu m/s$ and $U = 1600 \mu m/s$ are
%axisymmetric. In our simulations these vesicles are slightly asymmetric,
%are undergoing a tank treading motion, and we therefore report them as
%asymmetric slippers.
%
Using several initial reduced areas and flow rates, we recreate the
impermeable vesicle phase diagram for vesicles with $\beta = 10^{-3}$.
The equilibrium vesicle shape for different $U$ and initial reduced area
is summarized in Figure~\ref{fig:parabolicComposite}(a), and the
vertical displacement of the vesicle's center of mass is in
Figure~\ref{fig:parabolicComposite}(b). The effect of semipermeability
of a vesicle with $\alpha(0) = 0.9$ and $U=800\; \mu$m/s on its reduced
area and migration pattern are shown in
Figures~\ref{fig:parabolicComposite}(c)
and~\ref{fig:parabolicComposite}(d), respectively. 

 \begin{figure*}[hbp]
  \centering
  \includegraphics{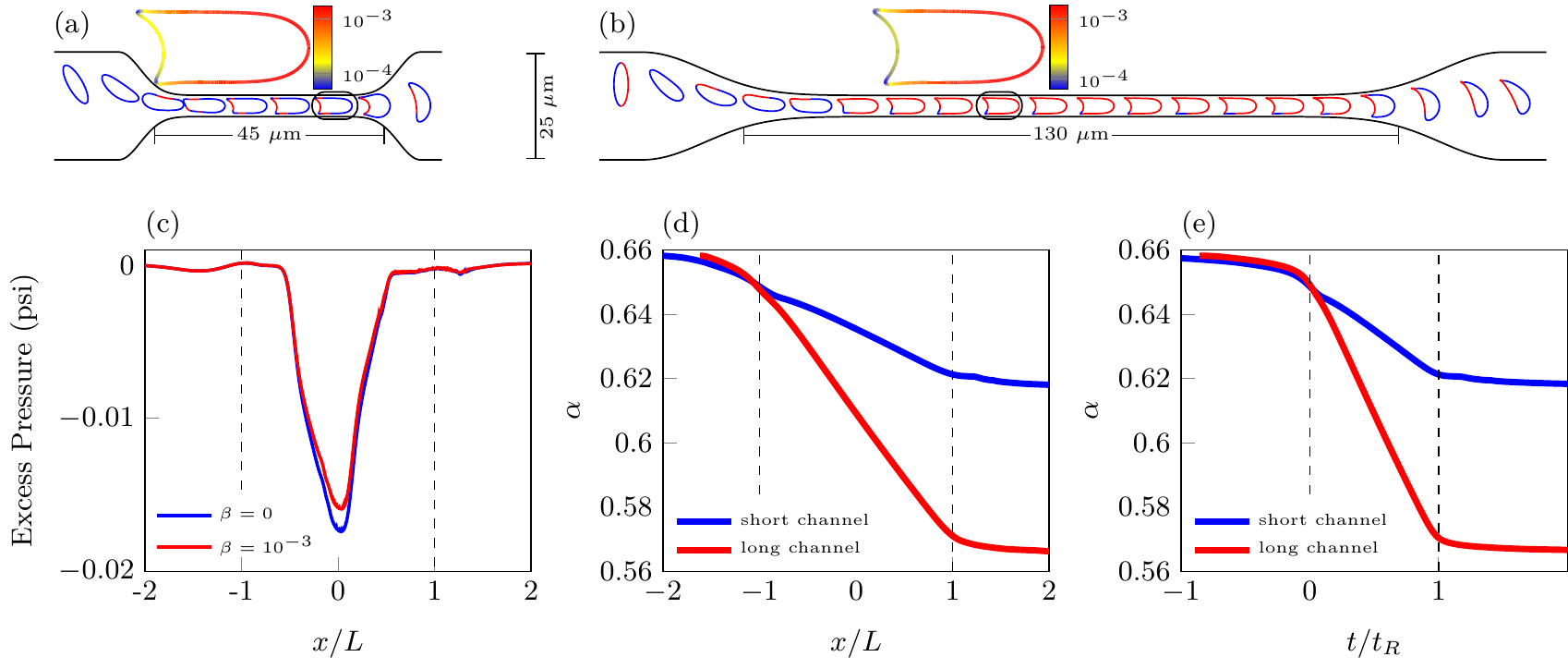}
  \caption{\label{fig:stenosisComposite} (a) A semipermeable vesicle
   passing through a closely-fit channel similar to the experimental
   device in~\cite{abk-fai-sto2006}. (b) A semipermeable vesicle passing
   through a  (four-times) longer closely-fit channel than (a). (c) The
   excess pressure of an impermeable and semipermeable vesicle in the
   short geometry in (a). (d) The reduced area of the semipermeable
   vesicles in the geometries in (a) and (b) as a function of the
   vesicles' center of mass. In (c) and (d), the x-axis is scaled by
   $1/L$ where $L$ is half the length of the constriction. (e) The
   reduced area of the semipermeable vesicles in the geometries in (a)
   and (b) as a function of time. The time axis is scaled by $1/t_R$
   where $t_R$ is the residency time. In parts (c) and (d), the dashed
   lines correspond to the locations of the inlet and outlet. In part
   (e), the dashed lines correspond to the times that the vesicle enters
   and exits the constriction. The colored vesicles show the tension
   distribution in N/m of the circled vesicles. Throughout the entire
   simulation, these vesicle configurations have the largest tension.}
\end{figure*}

As observed for the planar shear flow, the equilibrium reduced area of a
semipermeable vesicle in a Poiseuille flow is independent of the initial
reduced area and $\beta$, and  larger maximum flow velocity $U$ results
in smaller equilibrium reduced area
(Figure~\ref{fig:parabolicComposite}(a)). At the smallest velocity $U=
200\; \mu$m/s, the steady state reduced area is large, and the
equilibrium vesicle shape is an axisymmetric bullet. The equilibrium
reduced area decreases with increasing $U$, and the nonlinear flow
profile gives rise to an asymmetric tank-treading slipper for $U\ge 400$
$\mu m/s$. We observe that the membrane permeability to water
drastically alters the equilibrium vesicle shape and position relative
to the Poiseuille flow: At $\beta=10^{-3}$ the equilibrium vesicle
depends only on the maximum velocity $U$ of the nonlinear shear flow.
The higher $U$ the smaller the equilibrium reduced area and the farther
away the vesicle is relative to the flow center at $y=0$.  

%%%%%%%%%%%%%%%%%%%%%%%%%%%%%%%%%%%%%%%%%%%%%%%%%%%%%%%%%%%%%%%%%%%%%%%%
\section*{A semipermeable vesicle under strong confinement}
Here we consider a semipermeable vesicle under strong confinement: a
long closely-fit channel and a contracting channel. In a narrow
microfluidic channel, a large pressure jump is often needed to push the
vesicle through a closely-fit channel~\cite{abk-fai-sto2006}.  Our
simulations show that such a configuration can lead to an amplification
of water permeation at a time scale shorter than a few minutes. In
addition, we find that a large value of $\beta$ is needed to keep the
membrane tension below the threshold value for membrane poration. Such
high permeability for water may be regarded as an indication that the
lipid bilayer membrane is porated under strong confinement, and our
model vesicle is for a porated vesicle within the Helfrich free energy
framework.  As in the previous examples, we fix the vesicle length and
examine the enclosed water content as a dynamic consequence of water
flow, vesicle shape deformation, and membrane tension distribution.

%%%%%%%%%%%%%%%%%%%%%%%%%%%%%%%%%%%%%%%%%%%%%%%%%%%%%%%%%%%%%%%%%%%%%%%%
\subsection*{Semipermeable Vesicle in a Closely-fit Channel (Stenosis)}
When red blood cells go through small capillary vessels, they experience
large shape deformation and significant membrane stretching that
triggers ATP release~\cite{Wan2008_PNAS, ForsythWan2011_PNAS}.  Many
cells exhibit extraordinary flexibility as they go through narrow
vessels~\cite{AuStoreyMoore2016_PNAS}, and adjustment of
surface-to-volume ratio may be a key factor for a successful passage.
Here we quantify the effect of water permeability on a vesicle going
through a stenosis (a long microfluidic channel with a width smaller
than the cell radius).  Based on the findings in~\cite{abk-fai-sto2006},
we first examine the effect of semipermeability on the pressure
fluctuation across a stenosed geometry. In this experiment, a pressure
drop is used to squeeze a red blood cell through a constriction with
diameter $5\;\mu$m and length $45\;\mu$m. We set up a similar
computational geometry with a Poiseuille flow imposed at the inlet and
outlet (Figure~\ref{fig:stenosisComposite}(a)). We set the maximum flow
velocity to $U_{\max} = 1000 \;\mu$m/s at the inlet and outlet so that
the residency time of the vesicle in the stenosis channel is $t_R
\approx 25$ ms which agrees with the experimental
results~\cite{abk-fai-sto2006}. When the red blood cell enters the
channel, an increase in the pressure drop is required to maintain a
constant flow rate, and the difference in the pressure drop is called
the excess pressure.
\begin{figure}[htp]
  \centering
  \includegraphics{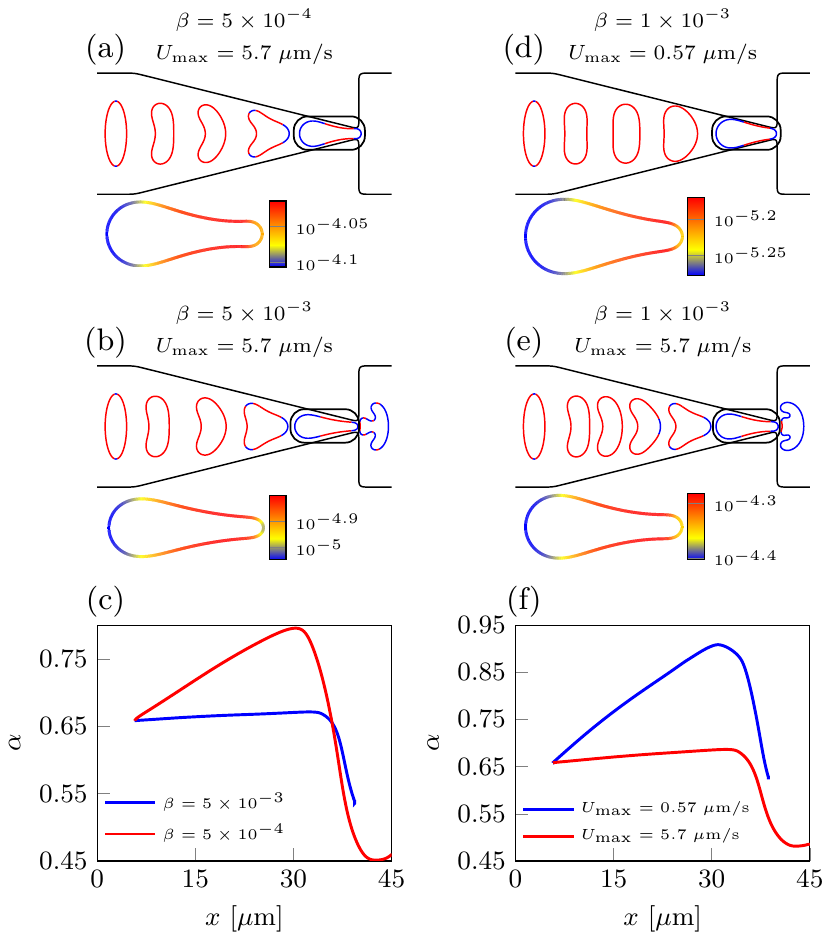}
  \caption{\label{fig:contractingComposite} A contracting channel
  similar to the experimental configuration in~\cite{wu2015critical}.
  The channel width varies from $18\;\mu$m to $1.8\;\mu$m, and it
  repeats every $45\;\mu$m. The fluid flux along the red regions is into
  the vesicle, and the fluid flux along the blue regions is out of the
  vesicle.  (a--c): The ability for a semipermeable vesicle to pass
  through the contraction depends on the permeability $\beta$. (d--f):
  The ability for a semipermeable vesicle to pass through the
  contraction depends on the flow rate. The colored vesicles show the
  tension distribution in N/m of the circled vesicles. Throughout the
  entire simulation, these vesicle configurations have the largest
  tension.}
\end{figure}

We consider both an impermeable and semipermeable vesicle with initial
reduced area $\alpha = 0.65$ passing through a long narrow channel and
compute the excess pressure as a function of vesicle position in the
channel (Figure~\ref{fig:stenosisComposite}(c)). The impermeable case
agrees with experimental results~\cite{abk-fai-sto2006} (see Figure 3).
Our simulations show that less excess pressure is needed to drive a
semipermeable vesicle through the closely-fit channel. Once in the
channel, the vesicle takes a bullet shape with maximum tension in the
front and minimum tension in the back
(Figure~\ref{fig:stenosisComposite}(a)), consistent with earlier
results~\cite{Pak2015_PNAS, HarmanBertrandJoos2017_CJP}.

Figures~\ref{fig:stenosisComposite}(d)
and~\ref{fig:stenosisComposite}(e) show that the vesicle deflates from
the beginning due to the high flow velocity, consistent with the results
for a vesicle in a Poiseuille flow in open space. As the vesicle enters
the stenosis channel, it continues to deflate in the front even though
we observe influx (red segment) at various rear locations in the
membrane. Overall, the vesicle loses about 6.1\% of its initial total
water at a constant rate as shown in
Figures~\ref{fig:stenosisComposite}(d)
and~\ref{fig:stenosisComposite}(e) (blue curves). Thus we expect that a
larger amount of deflation is possible if the vesicle is inside a longer
stenosis: we consider a vesicle passing through a channel of triple the
length in Figure~\ref{fig:stenosisComposite}(b) with
$U_{\max}=1000\;\mu$m/s, and we observe that vesicle stays inside the
stenosis for a residence time $t_R \approx 70$ ms and the area decreases
by 13.2\% (red curves).

%%%%%%%%%%%%%%%%%%%%%%%%%%%%%%%%%%%%%%%%%%%%%%%%%%%%%%%%%%%%%%%%%%%%%%%%
\subsection*{Semipermeable Vesicle in a Contracting Geometry} 
Squeezing a cell or an elastic vesicle through a slit in microfluidics
has gained popularity for its application in stress-induced release of
macromolecules~\cite{ShareiEtAl2013_PNAS, Pak2015_PNAS,
ZhangShenHoganBarakatMisbah2018_BJ, LuoBai2019_PoF}. Red blood cells go
through submicron slits many times during their lifetime, and this
process selects healthy, flexible red blood cells to
survive~\cite{wu2015critical,LuPeng2019_PoF}. Here we investigate the
role of membrane permeability to water when a semipermeable vesicle is
squeezed through a contracting microfluidic channel similar
to~\cite{wu2015critical}, where the width of the channel gradually
decreases from $18 \;\mu$m to $1.8\; \mu$m. The channel then immediately
reopens to $18\;\mu$m (Figure~\ref{fig:contractingComposite}). A
Poiseuille flow is imposed at the inlet and outlet with a constant
maximum flow rate of $U_{\max}$. We initialize a semipermeable vesicle
with $\alpha = 0.65$ that spans $10\; \mu$m of the total channel width
and investigate the effects of the membrane permeability to water and
the maximum flow velocity on whether a vesicle can successfully pass
through the slit. We color the vesicle according to the sign of its
flux---red for influx and blue for efflux.

In Figures~\ref{fig:contractingComposite}(a)
and~\ref{fig:contractingComposite}(b), we plot snapshots of vesicles
with different permeability $\beta$.
Figure~\ref{fig:contractingComposite}(c) shows the corresponding reduced
area as a function of the vesicle's location. We observe that if the
permeability is sufficiently large, the vesicle inflates gradually over
seconds before it reaches the neck, and then it quickly deflates within
a fraction of a second. However, if the permeability $\beta$ is too
low, the vesicle is too large when it reaches the narrowest part of the
channel, and it is unable to pass. In
Figures~\ref{fig:contractingComposite}(d)
and~\ref{fig:contractingComposite}(e), we plot snapshots of vesicles
with different imposed maximum velocities, and
Figure~\ref{fig:contractingComposite}(f) shows the reduced area as a
function of the vesicle's location. At low velocities, the vesicle
inflates similar to the unbounded parabolic flow example, and it is
unable to pass through the contraction. However, higher flow rates
result in additional deflation and the vesicle passes through the
contraction.

\begin{figure}[hbp]
  \centering
  \includegraphics{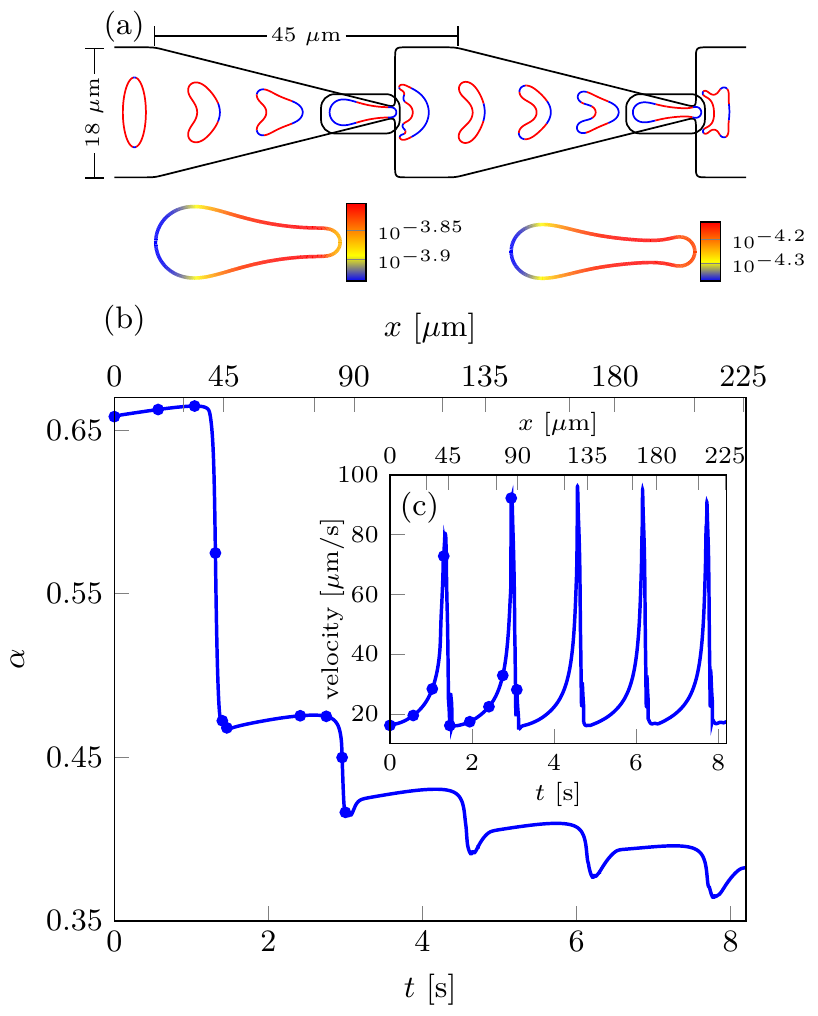}
  \caption{\label{fig:contractingComposite2} Vesicle reduced area and
  velocity vs.~time. The permeability coefficient is $\beta = 10^{-3}$
  and the maximum velocity is $18\;\mu$m/s. (a) A semipermeable vesicle
  passing through six periods of the contracting geometry. (b) The
  reduced area with the marks representing shapes in (a). (c) The
  vesicle velocity with the marks representing shapes in (a). In (b) and
  (c), the lower x-axis is the x-coordinate of the center of mass, and
  the upper x-axis is time. The colored vesicles show the tension
  distribution in N/m of the circled vesicles. Throughout the entire
  simulation, these vesicle configurations have the largest tension.}
\end{figure}

To understand the effects of water permeability on a vesicle that
repeatedly passes through micro-capillary vessels or sub-micron slits,
we construct a contracting geometry
(Figure~\ref{fig:contractingComposite2}) to simulate a vesicle passing
through the contracting geometry six times.  In
Figure~\ref{fig:contractingComposite2}(b), we show that the vesicle
deflates each time it passes through the contraction followed by a
gradual inflation (over seconds). The total area loss over the history
of the vesicle is almost 50\%. In
Figure~\ref{fig:contractingComposite2}(c), we show the vesicle velocity
as a function of the vesicle's center of mass. As the vesicle passes
through the contraction, it first accelerates until it reaches the neck
where it abruptly decelerates to go through the contraction. The reduced
area and velocity with respect to time are also shown in
Figure~\ref{fig:contractingComposite2}.

%%%%%%%%%%%%%%%%%%%%%%%%%%%%%%%%%%%%%%%%%%%%%%%%%%%%%%%%%%%%%%%%%%%%%%%%
\section*{Discussion}
In this work we investigate the effects of membrane permeability to
water on vesicle hydrodynamics. Without any external flow or
confinement, our analysis shows that a semipermeable vesicle always
inflates to maximize the water content inside the vesicle of a fixed
length (area) in two (three) dimensions. Results from analysis and
numerical simulation show that such relaxation of a semipermeable
vesicle is characterized by a rescaled time $\beta t$. We further
illustrate that the relaxation process consists of at least two phases:
The first phase is dominated by the balance between membrane traction
and viscous stress at early times (up to five minutes), when the vesicle
behaves like an impermeable membrane. The second phase sets in after a
few hours, when water influx begins to dominate and inflate the vesicle.
Over a time scale of more than four weeks, a semipermeable vesicle is
fully inflated, independent of its initial surface-to-volume ratio.

Under strong confinement we show that water permeation can be amplified
on a time scale of a few seconds as a vesicle is squeezed through a
closely-fit channel and a contracting channel. As a semipermeable
vesicle goes through a long closely-fit channel, we observe that the
vesicle loses water at a constant rate, and the longer the channel the
more the loss of water. In the case of a contracting channel, we find
that both permeability and the pressure gradient across the slit must be
sufficiently large for a successful passage. Throughout these
simulations the membrane tension is below the threshold for membrane
poration. Thus we expect the continuum membrane theory to well-capture
the vesicle hydrodynamics, and our simulation results shed light on the
configuration and time scales needed to observe the effects of
permeability on vesicle hydrodynamics under confinement. 

%Using scaling arguments, we showed that permeability due to osmolarity
%is two orders of magnitude larger than permeability due to tension, and
%six orders of magnitude larger than permeability due to bending.
%Therefore, future studies will investigate semipermeable membranes
%suspended in a solvent with a solute concentration. The presence of a
%solute is responsible for many physiological processes including cell
%migration in confined
%microenvironments~\cite{StrokaJiangChenEtAl2014_Cell,
%papadopoulos2008aquaporins}, dialysis~\cite{wan2006}, cell
%desiccation~\cite{vogl2014effect}, and
%apoptosis~\cite{marchetti1996mitochondrial}. 

Cells exhibit a dynamic volume-to-surface ratio in 
many physiological processes, such as 
cell migration in confined microenvironments~\cite{StrokaJiangChenEtAl2014_Cell, papadopoulos2008aquaporins}, 
dialysis~\cite{wan2006}, cell desiccation~\cite{vogl2014effect}, cell division~\cite{OdermattMiettinenKangEtAl2020_bioRxiv}, and
apoptosis~\cite{marchetti1996mitochondrial}.
%osmotic flow across the membrane.
While osmotic stress  is more effective and efficient than mechanical stress in contributing to change in cellular volume, our results suggest that the water permeation due to the capillary pressure may be non-negligible over long times or under strong confinement in simple configurations. 
To examine if this is the case in the physiological processes,  currently we are examining the combined effects of osmotic and mechanical stresses on water permeation in the small Peclet number limit, which is relevant to cellular migration in the presence of both confinement and a solute gradient~\cite{StrokaJiangChenEtAl2014_Cell} and has been studied numerically by Jaeger {\em et al.}~\cite{jae-car-med-try1999}.

%While the solute dynamics
%is governed by an advection-diffusion equation, many biophysical
%applications involve speeds no greater than $\sim 50$
%$\mu$m/hour~\cite{StrokaJiangChenEtAl2014_Cell}, a length scale of 10
%$\mu$m, and a solute diffusivity of $\sim 10^{-10}$ m$^2$/s (glucose in
%water). The resulting Peclet number is $\sim 10^{-2}$, and this limit
%has been analyzed for spherical vesicles by Anderson~\cite{and1983} and
%more general shapes using a finite element method by Jaeger {\em et
%al.}~\cite{jae-car-med-try1999}. Since the dynamics of the concentration
%is diffusion-dominated, we will use a stable and accurate boundary
%integral method that is similar to the numerical method we apply to the
%hydrodynamics (see Supplementary Material).

\acknow{We thank Howard Stone, Michael J.~Shelley, Petia Vlahovska,
Chaoqui Misbah, Manouk Abkarian, Sangwoo Shin, and Shravan
K.~Veerapaneni for discussions. B.Q.~acknowledges support from the
Simons Foundation, Mathematics and Physical Sciences-Collaboration
Grants for Mathematicians, Award Number 527139.  Y.-N.Y.~acknowledges
support from NSF (DMS 1614863, DMS 195160) and Flatiron Institute, part
of Simons Foundation.}

\showacknow{} % display acknowledgments

% \bibliography{refs}

%%%%%%%%%%%%%%%%%%%%%%%%%%%%%%%%%%%%%%%%%%%%%%%%%%%%%%%%%%%%%%%%%%%%%%%%
\section*{Supplementary Material: Formulation}
We consider a two-dimensional inextensible semipermeable vesicle
membrane $\gamma$ suspended in a viscous Newtonian fluid in domain
$\Omega$.  The vesicle membrane $\gamma$ is parameterized as $\xx(s,t)$
where $s$ is arclength and $t$ is time. In the suspending fluid the flow
velocity $\uu$ and pressure $p$ are governed by the incompressible
Stokes equations
\begin{align}
  \label{eqn:governing}
  -\nabla p + \mu \triangle \uu = 0, \quad
  \nabla \cdot \uu = 0, \qquad \xx \in \Omega \setminus \gamma,
\end{align}
where $\mu$ is the fluid viscosity. When $\Omega$ is an unconfined
geometry, the condition $\uu(\xx) \rightarrow \uu_\infty(\xx)$ as $|\xx|
\rightarrow \infty$ is enforced. In the confined examples, we enforce a
no-slip boundary condition along the top and bottom of the channel and a
Hagen–Poiseuille flow at the inlet and outlet---this boundary condition
has been used in similar vesicle setups~\cite{qua-bir2014,
rah-vee-bir2010, lu-rah-zor2017}. Along the vesicle membrane $\gamma$,
mass continuity, force balance, and local inextensibility are enforced
by requiring
\begin{align}
  \label{eqn:jump}
  \jump{\uu} = 0, \quad
  \jump{T\cdot\nn} = \ff_\mathrm{mem}, \quad
  \xx_s \cdot \uu_s = 0,
\end{align}
where $\nn$ is the outward normal of $\gamma$, $\jump{\cdot}$ is the
jump across the interface, and $T$ is the hydrodynamic stress tensor.
The membrane force $\ff_\mathrm{mem}$ is the sum of a bending force
with bending modulus $k_b$ and a force due to the tension $\Lambda$.
Following~\cite{vee-gue-zor-bir2009}, we introduce the variable $\sigma
= \Lambda - \frac{3}{2}\kappa^2$, where $\kappa$ is the vesicle curvature. Then,
the bending and tension forces corresponding to the Helfrich energy are
$\ff_\mathrm{ben} = -k_b \xx_{ssss}$ and $\ff_\mathrm{ten} = (\sigma
\xx_s)_s$. 

Semipermeability is introduced by modifying the vesicle velocity from
the standard no-slip condition $\dot{\xx} = \uu(\xx)$. Instead,
the permeating water flux through the lipid bilayer membrane is proportional to the
membrane traction~\cite{yao-mor2017}. That is, 
\begin{align}
  \label{eqn:vesicleFlux}
  \uu - \dot{\xx} = - k_w (\ff_\mathrm{mem} \cdot \nn) \nn, \qquad
  \xx \in \gamma,
\end{align}
where $k_w$ is the hydraulic conductivity.

%The governing equations are non-dimensionalized by introducing a length
%scale that we take to be $R_0 = 10^{-6}$m. Using a bending stiffness of
%$k_b = 10^{-19}$J and fluid viscosity $5 \times 10^{-2}$kg/ms, we define
%the time scale $\tau = \mu R_0^3/k_b = 0.5s$, the velocity scale $U =
%R_0/\tau = 2\mu$m/s, the pressure scale $P = k_b/R_0^3 = 10^{-1}$Pa, the
%tension scale $Q = k_b/R_0^2 = 10^{-7}$N/m, and the permeability scale
%$W = R_0\mu = 2 \times 10^{-5}$m$^2$s/kg. Another standard dimensionless
%variable in vesicle suspensions is the reduced area $\alpha = 4\pi A/L^2
%\in (0,1]$, where $L$ is the vesicle length and $A$ is its area. The
%length is conserved by the local inextensibility condition, but the area
%is not because of semipermeability. Therefore, the reduced area is
%dynamic.

The governing equations are non-dimensionalized by using a
characteristic length scale $R_0 = 10^{-6}$ m, a bending stiffness $k_b
= 10^{-19}$ J, and fluid viscosity $5 \times 10^{-2}$ kg/ms. We scale
time by $\tau = \mu R_0^3/k_b = 0.5\;s$, the velocity by $R_0/\tau =
2\;\mu$m/s, the pressure by $k_b/R_0^3 = 10^{-1}$ Pa, the tension by
$k_b/R_0^2 = 10^{-7}$ N/m, and the permeability by $R_0/\mu = 10^{-3}$
m$^2$s/kg. Under this nondimensionalization, a two-dimensional
semipermeable vesicle in a viscous fluid is characterized by (a) its
reduced area $\alpha = 4\pi A/L^2 \in (0,1]$, where $L$ is the vesicle
length and $A$ is its area, and (b) its permeability $\beta = k_w \mu
/R_0$. In two dimensions, the membrane length is conserved by the local
inextensibility condition, but the enclosed area is not conserved
because of semipermeability. Therefore, the reduced area is dynamic. We
convert the apparent permeability of a polyunsaturated PC
bilayer~\cite{OlbrichRawiczNeedhamEtAl2000_BJ} to the hydraulic
conductivity $k_w$ using the conversion
in~\cite{FettiplaceHaydon1980_PhysRev}, and found that $10^{-14}\le
k_w\le 10^{-12}$ m$^2$s/kg, thus $\beta$ is in the range $10^{-11}\le
\beta\le 10^{-9}$.

Since the fluid satisfies the Stokes equations, the velocity and
pressure can be represented as layer potentials. Given the interfacial
boundary conditions~\eqref{eqn:jump}, the fluid velocity is
\begin{align}
  \uu(\xx) = \uu_\infty(\xx) + \SS[\ff_\mathrm{mem}](\xx), \quad
    \xx \in \Omega,
\end{align}
where $\SS$ is the single-layer potential
\begin{align}
  \SS[\ff](\xx) = \frac{1}{4\pi} \int_{\gamma} \left(
    -\mathbf{I} \log\rho + \frac{\rr \otimes \rr}{\rho^2} \right)
    \ff(\yy) ds_{\yy},
%  \SS[\ff](\xx) = \frac{1}{4\pi\mu} \int_{\gamma} \left(
%    -\mathbf{I} \log\rho + \frac{\rr \otimes \rr}{\rho^2} \right)
%    \ff(\yy) ds_{\yy},
\end{align}
with $\rr = \xx - \yy$ and $\rho = |\rr|$. Imposing the flux
condition~\eqref{eqn:vesicleFlux} and the inextensibility condition, the
vesicle velocity satisfies the boundary integral equation
\begin{align}
  \label{eqn:vesVelocity}
  \dot{\xx} &= \uu_\infty(\xx) + \beta (\ff_\mathrm{mem}\cdot\nn)\nn
  + \SS [\ff_\mathrm{mem}](\xx),  \quad
  \xx_s \cdot \dot{\xx}_s = 0.
\end{align}
In confined geometries, the far field term $\uu_\infty(\xx)$ is replaced
by a double-layer potential with an unknown density defined on
$\partial\Omega$. The density function is used to satisfy the boundary
condition on $\partial\Omega$.

%%%%%%%%%%%%%%%%%%%%%%%%%%%%%%%%%%%%%%%%%%%%%%%%%%%%%%%%%%%%%%%%%%%%%%%%
\section*{Supplementary Material: Numerical Methods}
%%%%%%%%%%%%%%%%%%%%%%%%%%%%%%%%%%%%%%%%%%%%%%%%%%%%%%%%%%%%%%%%%%%%%%%%
\subsection*{Discretization in Space}
By using a boundary integral equation formulation, the unknowns are the
vesicle's position and tension and the unknown density function for
confined flows. Both $\gamma$ and $\partial\Omega$ are discretized at a
set of collocation points. Derivatives are computed with Fourier
differentiation, and the eighth-order quadrature~\cite{alp1999} is
applied to the weakly singular single-layer potential. For confined
geometries, the trapezoid rule is used to approximate the double-layer
potential since its kernel is smooth and periodic. Nearly-singular
integrals are computed with an interpolation-based quadrature
rule~\cite{qua-bir2014}.

%%%%%%%%%%%%%%%%%%%%%%%%%%%%%%%%%%%%%%%%%%%%%%%%%%%%%%%%%%%%%%%%%%%%%%%%
\subsection*{Discretization in Time}
To eliminate stringent time step restrictions, we modify a time stepping
method that discretizes high-order derivatives
semi-implicitly~\cite{vee-gue-zor-bir2009}. We introduce the notation
\begin{alignat}{3}
  &B[\xx]\ff = -\frac{d^4}{ds^4} \ff,  \qquad
  &&T[\xx]\sigma = (\sigma \xx_s)_s, \\
  &D[\xx]\ff = \xx_s \cdot \ff_s, 
  &&P[\xx]\ff = (\ff \cdot \nn) \nn,
\end{alignat}
for the differential and projection operators. Then, the
governing equations~\eqref{eqn:vesVelocity} are
\begin{align}
  \dot{\xx} &= \uu_\infty(\xx) + \beta P(B\xx + T\sigma)
  + \SS (B\xx + T\sigma), \quad
  D \dot{\xx} = 0.
\end{align}
We note that the integral and differential operators, which all depend
on the vesicle shape, are non-linear with respect to the vesicle
shape. We construct a time stepping scheme by letting $B^N$ be the
bending operator due to the vesicle configuration at time $t^N$, and use
similar notation for the other operators. Then, a first-order
semi-implicit time stepping method of~\eqref{eqn:vesVelocity} is
\begin{alignat}{3}  
  \frac{\xx^{N+1} - \xx^N}{\triangle t} &= \uu_\infty(\xx^N) 
  &&+ \beta P^N\left(B^N\xx^{N+1} + T^N\sigma^{N+1}\right) \nonumber \\
  & &&+ \SS^N\left(B^N\xx^{N+1} + T^N\sigma^{N+1}\right),  \\
  D^N\xx^{N+1} &= 1,
\end{alignat}
and this linear system is solved with matrix-free GMRES.

To achieve second-order accuracy over long time horizons, we modify an
adaptive spectral deferred correction time stepping
method~\cite{qua-bir2016}. Since the area of the vesicle is not
conserved, we only use the length of the vesicle to estimate the error
which determines if a time step is accepted or not and determines the
subsequent time step size.

\begin{figure}[htp]
  \centering
  \includegraphics{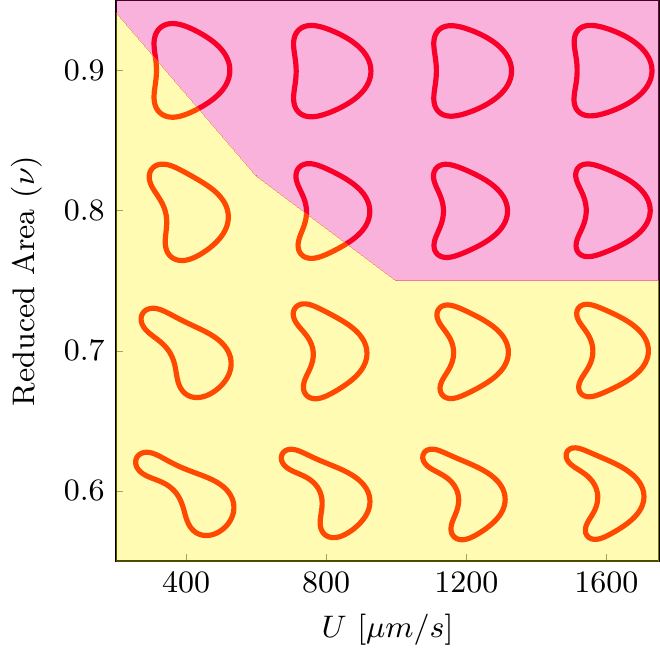}
  \caption{\label{fig:PoiseuillePhase} Phase diagram of an impermeable
  vesicle with various reduced areas submerged in a Poiseuille flow with
  various flow rates. The shapes agree qualitatively with Kaoui {\em et
  al.}~\cite{kao-bir-mis2009} (see Figure 2).}
\end{figure}

%%%%%%%%%%%%%%%%%%%%%%%%%%%%%%%%%%%%%%%%%%%%%%%%%%%%%%%%%%%%%%%%%%%%%%%%
\subsection*{Numerical Validation}
To validate our numerical codes for simulating vesicle hydrodynamics, we reproduce a phase diagram of the equilibrium
shape of an impermeable vesicle in a Poiseuille flow. Depending on
the maximum flow velocity and the vesicle reduced area, the equilibrium vesicle shape is either an
axisymmetric bullet, an axisymmetric parachute, or a tank-treading
asymmetric slipper~\cite{kao-bir-mis2009}. Comparing our results
(Figure~\ref{fig:PoiseuillePhase}) with those in~\cite{kao-bir-mis2009}
(see Figure 2), we observe qualitative agreement in the equilibrium
vesicle shapes except that,
%
%see that the vesicle shapes qualitatively agree. We
%note that 
%
Kaoui {\em et al.}~report that the vesicles with $\alpha=0.7$
and flow rates $U = 1200\; \mu$m/s and $U = 1600\; \mu$m/s are
axisymmetric.  In our simulations these vesicles are slightly
asymmetric, are undergoing a tank-treading motion, and we therefore
report them as asymmetric slippers.

%%%%%%%%%%%%%%%%%%%%%%%%%%%%%%%%%%%%%%%%%%%%%%%%%%%%%%%%%%%%%%%%%%%%%%%%
\section*{Supplementary Material: Vesicle Area}
In the absence of an imposed flow, the steady state vesicle shape and a
differential equation for the area of the vesicle can be derived. We
start by following the analysis of Veerapaneni {\em et
al.}~\cite{vee-raj-bir-pur2009} by computing the steady state shape of a
semipermeable vesicle in a quiescent flow. The steady state shape is a
stationary point of the Lagrangian
\begin{align}
  \mathcal{L} = \frac{1}{2}\int_{\gamma} \kappa^2 \, ds +
    \Lambda \left(\int_{\gamma} ds  - L \right),
\end{align}
where the Lagrange multiplier $\Lambda$ is the tension. Note the absence
of the pressure Lagrange multiplier since the total water content inside the vesicle membrane is not conserved.
Taking the variation of $\mathcal{L}$ with respect to $\gamma$ and
setting it to zero, the steady state shape satisfies $\kappa_{ss} +
\frac{1}{2}\kappa^3 - \Lambda \kappa = 0$, with initial conditions
$\kappa(0) = \kappa_0$ and $\kappa_s(0) = 0$.  Integrating once, we have
\begin{align}
  \frac{\kappa_s^2}{2} + \frac{\kappa^4}{8} - 
    \frac{\Lambda}{2}\kappa^2 = \frac{\kappa_0^4}{8} - 
    \frac{\Lambda}{2}\kappa_0^2.
  \label{eqn:curveODE}
\end{align}
Assuming $\kappa_s \neq 0$,~\eqref{eqn:curveODE} is separable, and can
be solved analytically in terms of the {\em EllipticF} function using
Mathematica. Unlike in the impermeable case, the solution
of~\eqref{eqn:curveODE} is not periodic unless $\kappa(s) = \kappa_0$.
Therefore, the equilibrium shape of a semipermeable vesicle in a
quiescent flow is a circle with radius $L/2\pi$ and constant tension
$\Lambda = \kappa^2/2 = 2\pi^2/L^2.$ Note that in contrast to a
impermeable vesicle, the steady state tension is positive rather than
negative. Such equilibrium shape of a semipermeable vesicle results from the balance between elastic stress and membrane tension. 
For an elastic membrane with a large pore, the equilibrium membrane (hemispherical) shape also results from such balance \cite{Ryham2018_JFM}.

We next compute the transient area of the vesicle. Since the area of the
vesicle is
\begin{align}
  A(t) = \frac{1}{2}\int_{\gamma} (\xx \cdot \nn)\, ds,
  \label{eqn:area}
\end{align}
its derivative includes terms due to the time derivative of $\xx \cdot
\nn$ and a term due to interface stretching~\cite{lai-tse-hua2008}, but
this latter term is zero because of the local inextensibility condition.
Therefore,
\begin{align}
  \dot{A}(t) =
  \frac{1}{2} \int_{\gamma} (\dot{\xx} \cdot \nn)\, ds  + 
  \frac{1}{2} \int_{\gamma} (\xx \cdot \dot{\nn})\, ds.
\end{align}
Applying integration by parts, the two integrals agree, and we have
\begin{align}
  \dot{A}(t) = \int_\gamma \left(\beta (\ff \cdot \nn)\nn 
    + \SS[\ff]\right) \cdot \nn\, ds 
  = \beta \int_\gamma (\ff \cdot \nn)\, ds,
\end{align}
where we have used the incompressibility of the single-layer potential.
Repeatedly applying integration by parts, we obtain
\begin{align}
  \dot{A}(t) & = \beta \int_\gamma \left(
    \frac{\kappa^3}{2} - \kappa \Lambda \right) \, ds.
  \label{eqn:areaROC}
\end{align}
%\begin{align}
%  \dot{A}(t) &= \beta \int_\gamma \left( -\xx_{ssss} + 
%    (\sigma \xx_s)_s \right) \cdot \nn \, ds \\
%  &= -\beta \int_\gamma \left(-\xx_{sss} + \sigma \xx_s 
%    \right) \cdot \kappa \xx_s \, ds \\
%  &= -\beta \int_\gamma \left(\xx_{ss} \cdot 
%    (\kappa \xx_s)_s + \kappa \sigma \right) \, ds \\
%  &= -\beta \int_\gamma \left(\kappa (\xx_{ss} \cdot \xx_{ss}) + 
%    \kappa_s (\xx_{ss} \cdot \xx_s) + \kappa \sigma \right) 
%    \, ds \\
%  &  = -\beta \int_\gamma \left(\kappa^3 + \kappa \sigma \right) 
%    \, ds = \beta \int_\gamma \left(
%    \frac{\kappa^3}{2} - \kappa \Lambda \right) \, ds.
%  \label{eqn:areaROC}
%\end{align}
Note that the integrand in~\eqref{eqn:areaROC} is zero when the membrane tension balances the elastic force:
\begin{align}
\label{eqn:areaROC2}
\Lambda &= \frac{\kappa^2}{2}=\frac{\kappa_0^2}{2} \;\mathrm{ at }\; \mathrm{equilibrium},
\end{align}
which is consistent with the equilibrium results in the vesicle area dynamics.

%reaches its steady state which confirms that $\dot{A}(t) = 0$.

%%%%%%%%%%%%%%%%%%%%%%%%%%%%%%%%%%%%%%%%%%%%%%%%%%%%%%%%%%%%%%%%%%%%%%%
\section*{Supplementary Material: Vesicle Tensions}
\begin{figure*}[htp]
  \centering
  \includegraphics[width=\textwidth]{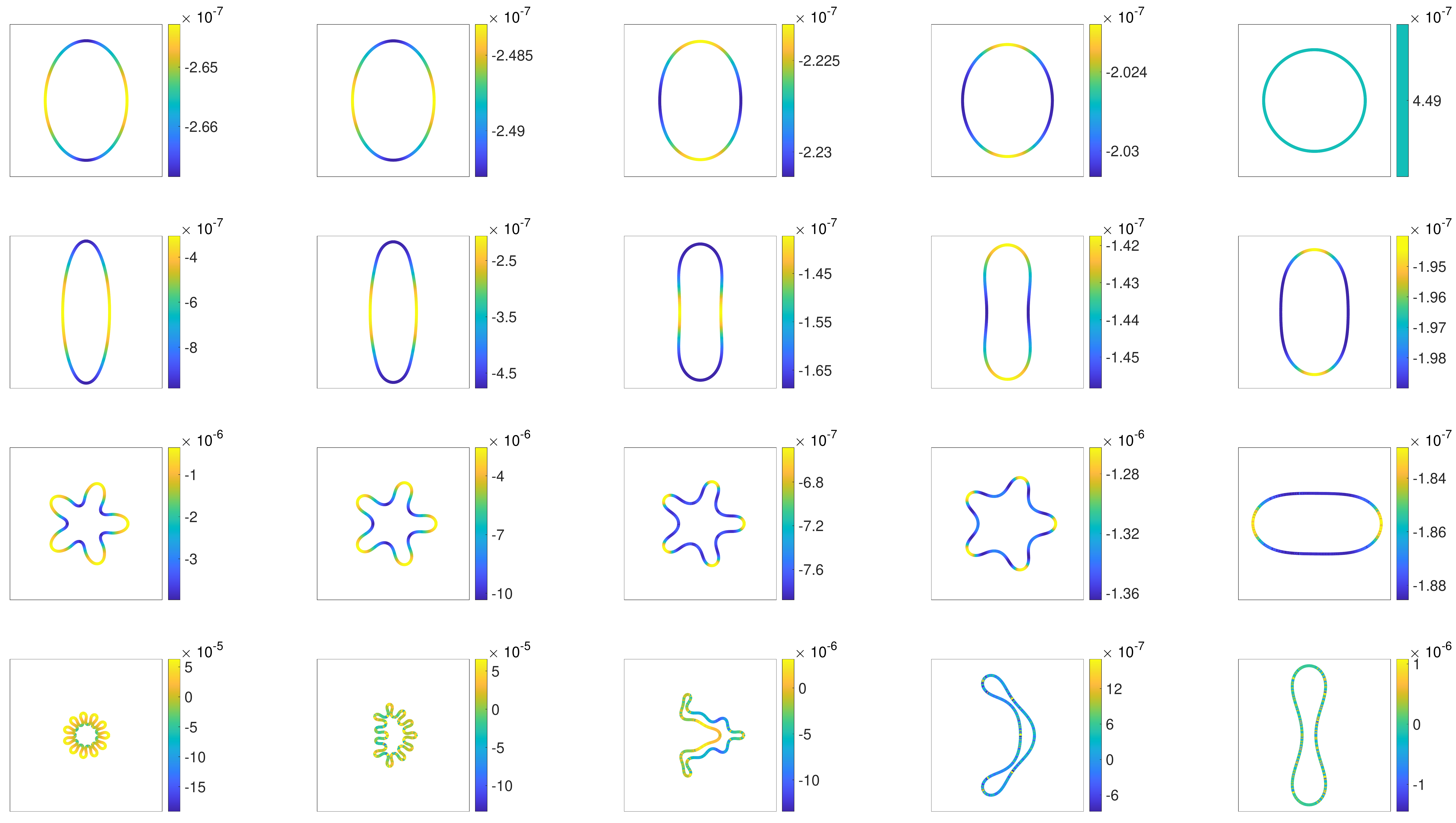}
  \caption{\label{fig:relaxationTensions} The tension distribution of
  vesicles in relaxation that corresponds to
  Figure~\ref{fig:relaxationComposite}.}
\end{figure*}

\subsection*{Vesicle Relaxation}
Figure~\ref{fig:relaxationTensions} shows the membrane tension
distribution along the relaxing vesicles in
Figure~\ref{fig:relaxationComposite}. The membrane tension magnitude
(color bars) is in N/m.  As the semipermeable vesicle relaxes, the
membrane tension reduces in magnitude and becomes more evenly
distributed along the membrane. In the top row ($\alpha(0)=0.95$) the
vesicle nearly relaxes to a circle on the right, and the membrane
tension is uniform and equal to $k_b \kappa_0^2/2$ as predicted by our
analysis (see~\eqref{eqn:areaROC2}). In the bottom row
($\alpha(0)=0.12$), the initial vesicle shape has a lot of high
curvature regions (and thus the low reduced area), and it takes more
than four hours to inflate to twice the initial total water content for
$\beta=10^{-7}$. In contrast the water enclosed inside the vesicle in
the top row remains nearly constant in the first four hours.

\subsection*{Vesicle in a Planar Shear Flow}
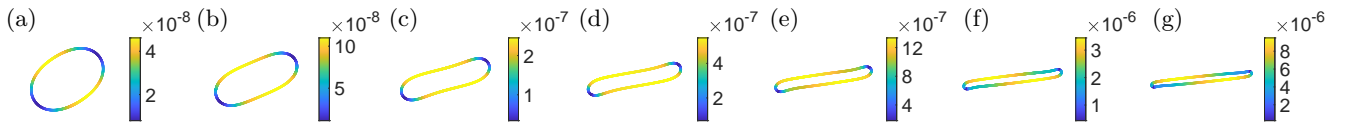
\begin{figure*}[hbp]
  \begin{tikzpicture}[scale=1]
    \node[inner sep=0pt] at (0,0) {
    \includegraphics[width=\textwidth]{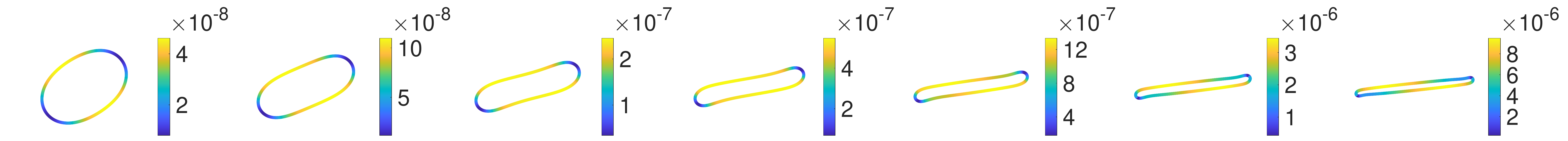}
    };
    \node at (-8.54,0.6) {(a)};
    \node at (-6.00,0.6) {(b)};
    \node at (-3.46,0.6) {(c)};
    \node at (-0.92,0.6) {(d)};
    \node at (+1.60,0.6) {(e)};
    \node at (+4.16,0.6) {(f)};
    \node at (+6.70,0.6) {(g)};
  \end{tikzpicture}
  \caption{\label{fig:shearTensions} The tension distribution along a
  vesicle in a linear shear flow. The semipermeability constant is
  $\beta = 10^{-4}$ for all these cases, and the elastic capillary
  number ranges from $10^{-1}$ to $10^{2}$ as in
  Figure~\ref{fig:shearComposite}.}
\end{figure*}

Our simulations of a semipermeable vesicle in a planar shear flow show
that the equilibrium vesicle shape depends mostly on the capillary
number $Ca_\mathrm{E}$, independent of the dimensionless water
permeability $\beta$. Our simulations show that the time it takes for a
semipermeable vesicle to reach equilibrium in a planar shear flow is
inversely proportional to $\beta$, just like the relaxing vesicle. The
tension distribution along the equilibrium vesicle at $\beta=10^{-4}$ is
shown in Figure~\ref{fig:shearTensions}, where the color bars are for
the membrane tension in N/m. The capillary number $Ca_\mathrm{E}$
increases from Figure~\ref{fig:shearTensions}(a) to (g), and we observe
that the tension magnitude also increases almost linearly with
$Ca_\mathrm{E}$. We also observe that the membrane tension is relatively
small at the tip when the membrane curvature is large due to the
extensional component in the shear flow.

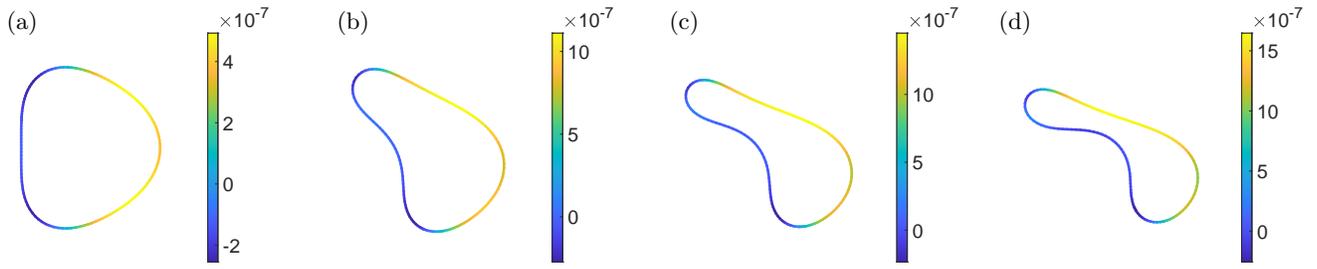
\begin{figure*}[htp]
  \centering
  \begin{tikzpicture}[scale=1]
    \node[inner sep=0pt] at (0,0) {
    \includegraphics[width=\textwidth]{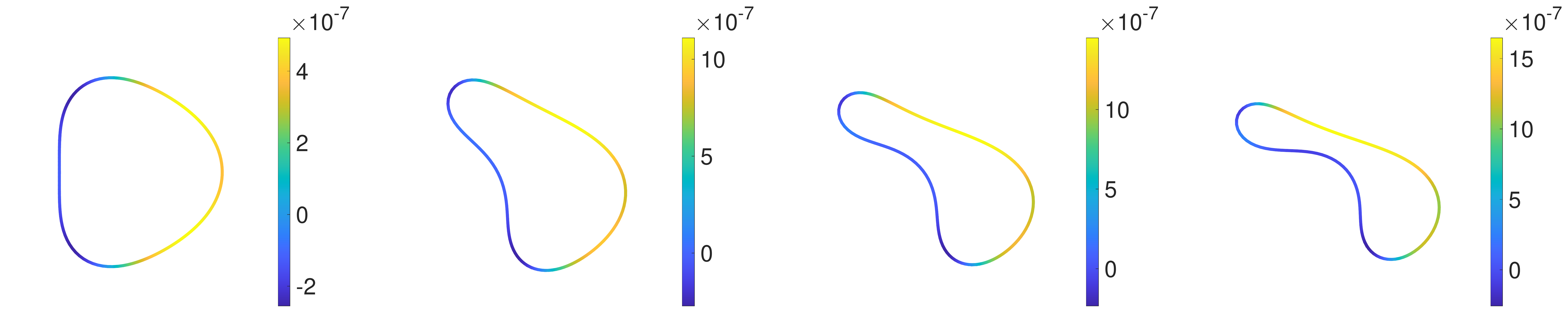}
    };
    \node at (-8.2,1.5) {(a)};
    \node at (-3.8,1.5) {(b)};
    \node at (+0.6,1.5) {(c)};
    \node at (+5.0,1.5) {(d)};
  \end{tikzpicture}
  \caption{\label{fig:parabolicTensions} The equilibrium tension
  distribution of a semipermeable vesicle with $\beta=10^{-3}$ in a parabolic
  flow (Figure~\ref{fig:parabolicComposite}). The flow velocities
  are (a) 200 $\mu$m/s, (b) 400 $\mu$m/s, (c) 600 $\mu$m/s, and (d)
  800 $\mu$m/s.}
\end{figure*}

\subsection*{Vesicle in a Poiseuille flow}
In a nonlinear shear flow, an impermeable vesicle can reach equilibrium
position at the mid-plane of the shear flow for sufficiently large
reduced area, regardless of the maximum flow velocity $U$. For a
semipermeable vesicle, however, our simulation results (summarized in
Figure~\ref{fig:parabolicComposite}) show that the equilibrium vesicle
shape and position are determined by $U$, and independent of the initial
vesicle reduced area and permeability $\beta$. The corresponding
membrane tension distribution is shown in
Figure~\ref{fig:parabolicTensions} as a function of flow velocity $U$
from $200\;\mu$m/s to $800 \;\mu$m/s.

We note that in the Poiseuille flow the membrane tension is positive in
the front, and negative in the rear part of the vesicle. Furthermore we
also note that the tension magnitude is almost linearly proportional to
$U$ as in the linear shear flow.

\subsection*{Vesicle in a Closely-Fit Channel}
\begin{figure*}[htp]
  \centering
  \begin{tikzpicture}[scale=1]
    \node[inner sep=0pt] at (0,0) {
    \includegraphics[width=\textwidth]{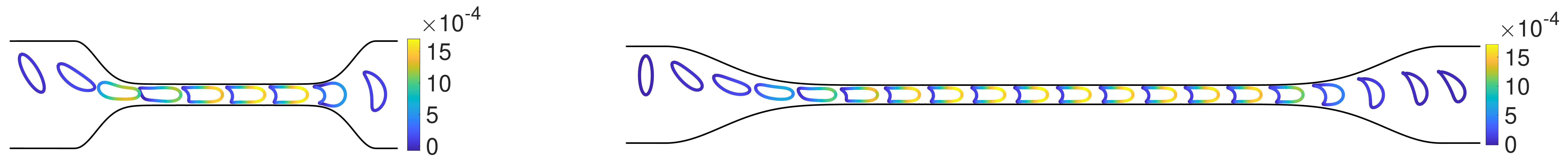}
    };
    \node at (-8.6,0.7) {(a)};
    \node at (-1.6,0.7) {(b)};
  \end{tikzpicture}
  \caption{\label{fig:stenosisTensions} The tension distribution along a
  semipermeable vesicle with $\beta = 10^{-3}$ going through a
  closely-fit channel of two different lengths.}
\end{figure*}

When a semipermeable vesicle goes through a closely-fit channel, we observe that the water efflux is found in the front, and the influx is in the rear of the vesicle membrane (Figure~\ref{fig:stenosisComposite}). This corresponds well to the corresponding tension distribution in Figure~\ref{fig:stenosisTensions}: Larger tension in the front, and smaller tension in the rear.

\subsection*{Vesicle Going Through a Contracting Channel (Slit)}
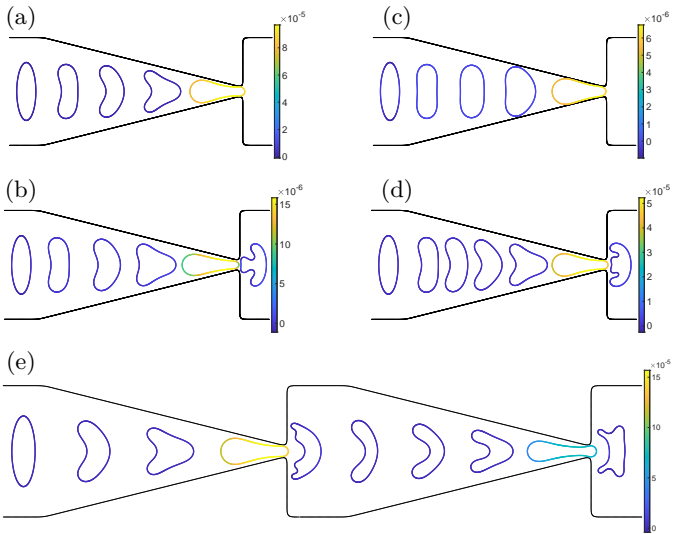
\begin{figure}[htp]
  \centering
  \begin{tikzpicture}[scale=1]
    \node[inner sep=0pt] at (0,0) {
    \includegraphics[width=0.5\textwidth]{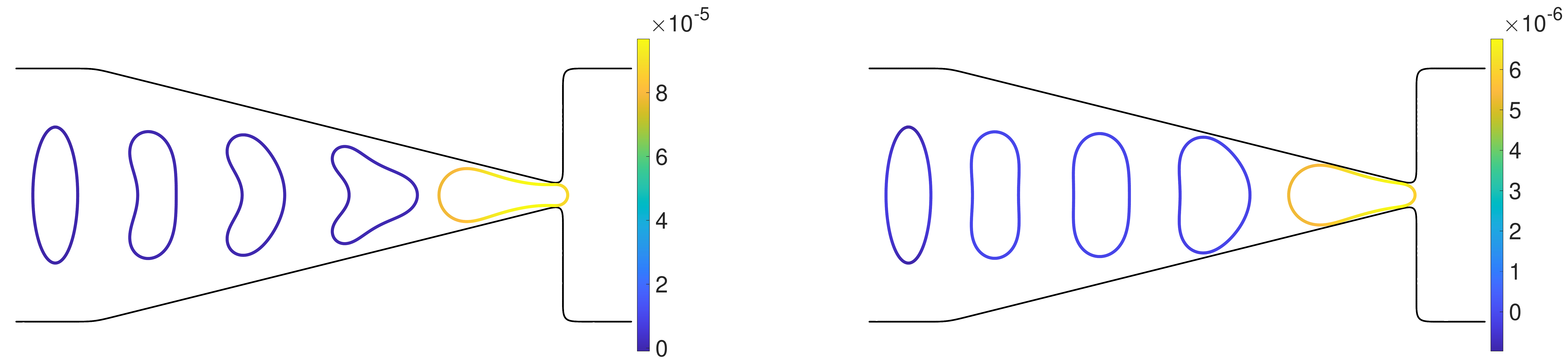}
    };
    \node[inner sep=0pt] at (0,-2.3) {
    \includegraphics[width=0.5\textwidth]{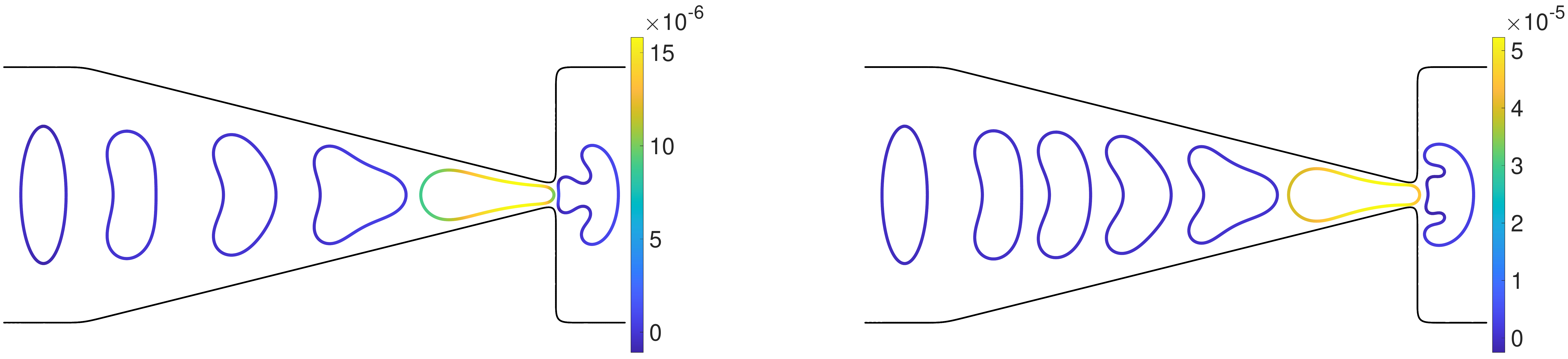}
    };
    \node[inner sep=0pt] at (0,-4.8) {
    \includegraphics[width=0.5\textwidth]{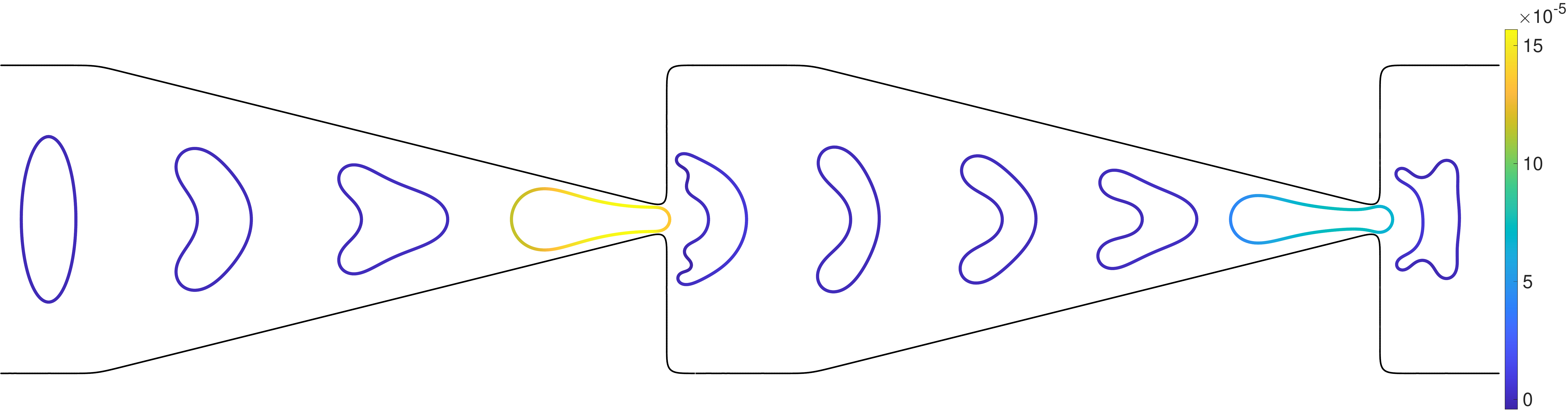}
    };
    \node at (-4.2,0.9) {(a)};
    \node at (-4.2,-1.4) {(b)};
    \node at (+0.8,0.9) {(c)};
    \node at (+0.8,-1.4) {(d)};
    \node at (-4.2,-3.7) {(e)};
  \end{tikzpicture}
  \caption{\label{fig:contractingTensions1} The tension distribution
  along a semipermeable vesicle going through a contracting geometry.
  The vesicle tension increases by up to five orders of magnitude when
  as it enters the neck. The permeability constant and flow rates are
  (a) $U_{\max} = 5.7\;\mu$m/s, $\beta = 5 \times 10^{-4}$; (b)
  $U_{\max} = 5.7\;\mu$m/s, $\beta = 5 \times 10^{-3}$; (c) $U_{\max} =
  0.57\;\mu$m/s, $\beta = 1 \times 10^{-3}$; (d) $U_{\max} =
  5.7\;\mu$m/s, $\beta = 1 \times 10^{-3}$; (e) $U_{\max} = 18\;\mu$m/s,
  $\beta = 1 \times 10^{-3}$.}
\end{figure}

When a semipermeable vesicle goes through a contracting channel (slit)
the numerical simulations show that both the driving flow and the
membrane permeability have to be sufficiently large for a successful
passage through the slit (Figure~\ref{fig:contractingComposite}). As
the vesicle goes through the contracting channel, the gradual
confinement first leads to a slow inflation. Once the vesicle is close
to the neck, it quickly deflates in order to fit the constriction. When
the vesicle is near the neck, simulations show that the water efflux is
in the front and rear of vesicle while most water influx is in the
middle. The tension distribution along the vesicle as a function of
their position in the contracting channel is summarized in
Figure~\ref{fig:contractingTensions1}, where we observe the
tension has the largest magnitude when the vesicle is at the neck. 

\end{document}

%% file: my_macros2.tex
\newcommand{\beq} {\begin{equation}}
\newcommand{\eeq} {\end{equation}}
\newcommand{\bit}{\begin{itemize}}
\newcommand{\eit}{\end{itemize}}
\newcommand{\bde}{\begin{description}}
\newcommand{\ede}{\end{description}}
\newcommand{\bce}{\begin{center}}
\newcommand{\ece}{\end{center}}
\newcommand{\ben} {\begin{enumerate}}
\newcommand{\een} {\end{enumerate}}
\newcommand{\bea} {\begin{eqnarray}}
\newcommand{\eea} {\end{eqnarray}}
\newcommand{\barr} {\begin{array}}
\newcommand{\earr} {\end{array}}
\newcommand{\bean} {\begin{eqnarray*}}
\newcommand{\eean} {\end{eqnarray*}}
\renewcommand{\SS}{\mathcal{S}}

\newcommand{\xx}{\mathbf{x}}
\newcommand{\yy}{\mathbf{y}}

\newcommand{\nn}{\mathbf{n}}

\newcommand{\rr}{\mathbf{r}}

\newcommand{\ff}{\mathbf{f}}
\newcommand{\uu}{\mathbf{u}}

\renewcommand{\ss}{\mathbf{s}}

\newcommand{\jump}[1]{\left[\!\left[ #1 \right]\!\right]_\gamma}

\usepackage{pgfplots}
\usetikzlibrary{matrix}
\usepgfplotslibrary{groupplots}
\usepgfplotslibrary{patchplots}
\usepackage{color}
\usepackage{array}
\usepackage{amsmath,amsfonts,bm}
\usepackage{graphicx,epsfig}
\usepackage{todonotes}

\newcolumntype{C}[1]{>{\centering\let\newline\\\arraybackslash\hspace{0pt}}m{#1}}